\documentclass[aps,pre, amsmath,amssymb,
review,%
]{revtex4-1}

\usepackage{graphicx}% Include figure files
\usepackage{dcolumn}% Align table columns on decimal point
\usepackage{bm}% bold math
\usepackage[mathlines]{lineno}% Enable numbering of text and display math

\usepackage[utf8]{inputenc}
\usepackage[T1]{fontenc}
\usepackage{times}
\usepackage{mathtools}
\usepackage{newtxmath}
\usepackage{booktabs, siunitx}
\usepackage[table,xcdraw]{xcolor}
\usepackage{hyperref}
\usepackage{float}
\usepackage{here}
\usepackage{subcaption}
\hypersetup{
    colorlinks =    true,
    linkcolor =     [rgb]{0.0,0.0,1.0},
    anchorcolor =   [rgb]{0.0,0.0,1.0},
    citecolor =     [rgb]{0,0,1},
    filecolor =     [rgb]{0.0,0.0,1.0},
    urlcolor =      [rgb]{0.0,0.0,1.0},
}
\usepackage[nameinlink]{cleveref}
\usepackage{multirow}
\newcolumntype{L}[1]{>{\raggedright\let\newline\\\arraybackslash\hspace{0pt}}m{#1}}
\newcolumntype{C}[1]{>{\centering\let\newline\\\arraybackslash\hspace{0pt}}m{#1}}
\newcolumntype{R}[1]{>{\raggedleft\let\newline\\\arraybackslash\hspace{0pt}}m{#1}}

 \crefname{figure}{Fig.}{Figs.}
 \crefname{equation}{Eq.}{Eqs.}
 \crefname{table}{Table}{Tables}
 \crefname{section}{Sec.}{Secs.}

 \crefname{figure}{Figure}{Figures}
 \crefname{equation}{Equation}{Equations}
 \crefname{table}{Table}{Tables}
\makeatletter
\newcommand{\doublecheck}{\textcolor{blue}{\checkmark}\kern-0.6em\textcolor{red}{\checkmark}}

\DeclarePairedDelimiter\abs{\lvert}{\rvert}%
\let\oldabs\abs
\def\abs{\@ifstar{\oldabs}{\oldabs*}}

\makeatother

\begin{document}

\preprint{Physical Review E}

\title{Voxelization based packing analysis for discrete element simulations of non-spherical particles}
% Force line breaks with \\

\author{Venkata Rama Manoj Pola}
\affiliation{Department of Mechanical Engineering, Indian Institute of Technology Madras, Chennai- 600036, India}%
\affiliation{Additive Manufacturing Group - Centre of Excellence in Materials and Manufacturing for Futuristic Mobility, Indian Institute of Technology Madras,Chennai - 600036,India}
\author{Raghuram Karthik Desu}
\affiliation{Department of Mechanical Engineering, National Institute of Technology, Tiruchirappalli - 620015, India}%
\author{Ratna Kumar Annabattula}%
\email[Corresponding author: ]{ratna@iitm.ac.in}
\affiliation{Department of Mechanical Engineering, Indian Institute of Technology Madras, Chennai- 600036, India}%
\affiliation{Additive Manufacturing Group - Centre of Excellence in Materials and Manufacturing for Futuristic Mobility, Indian Institute of Technology Madras,Chennai - 600036,India}

\date{\today}% It is always \today, today,

\begin{abstract}
A voxelization based post-processing algorithm is proposed to analyze the packing of non-spherical particle assemblies simulated using the Discrete Element Method. Voxelization of the particle data allows for isolating the geometric features of the granular assembly in various spatial sub-domains (2D surface or 3D region) and investigate the localized packing behaviour. Analyzing the local packing behaviour enables determining confined influences such as the wall-effect, stacking behaviour, local expansion/contraction and localized loading. The efficacy of the proposed technique to analyze practical granular assemblies is demonstrated through the packing analysis of different assemblies of superquadric cubes, superquadric ellipsoidals, and multi-spherical coffee beans. The python codes developed for the implementation of the algorithm and the results are provided at \url{https://github.com/Manojpvr/VoxelDEM}. 
\end{abstract}

\maketitle
%=================================================%
\section{Introduction}\label{Sec:Intrductionoduction}
\begin{figure}[htbp]
\begin{center}
\includegraphics[width=8.6 cm]{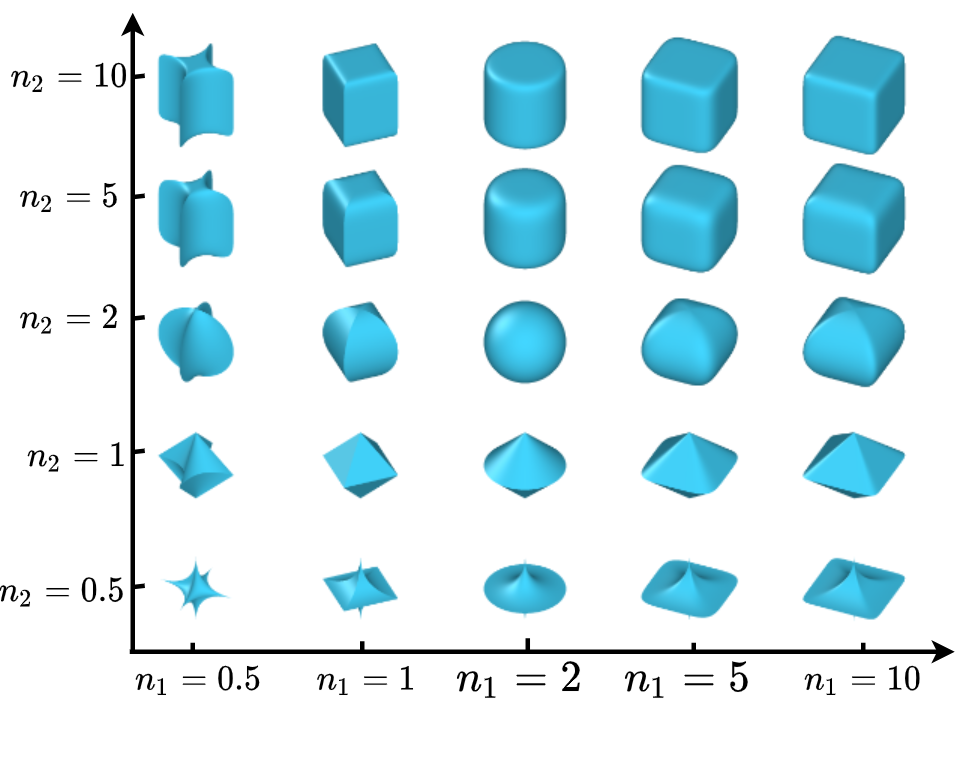}
\caption{Representation of the various shapes that can be represented by changing the shape parameters $(n_1,n_2)$ in \cref{sq_equation}}
\label{Fig: SQ_shapes}
\end{center}
\end{figure}

\begin{figure}[htbp]
\centering
\includegraphics[width=8.6 cm]{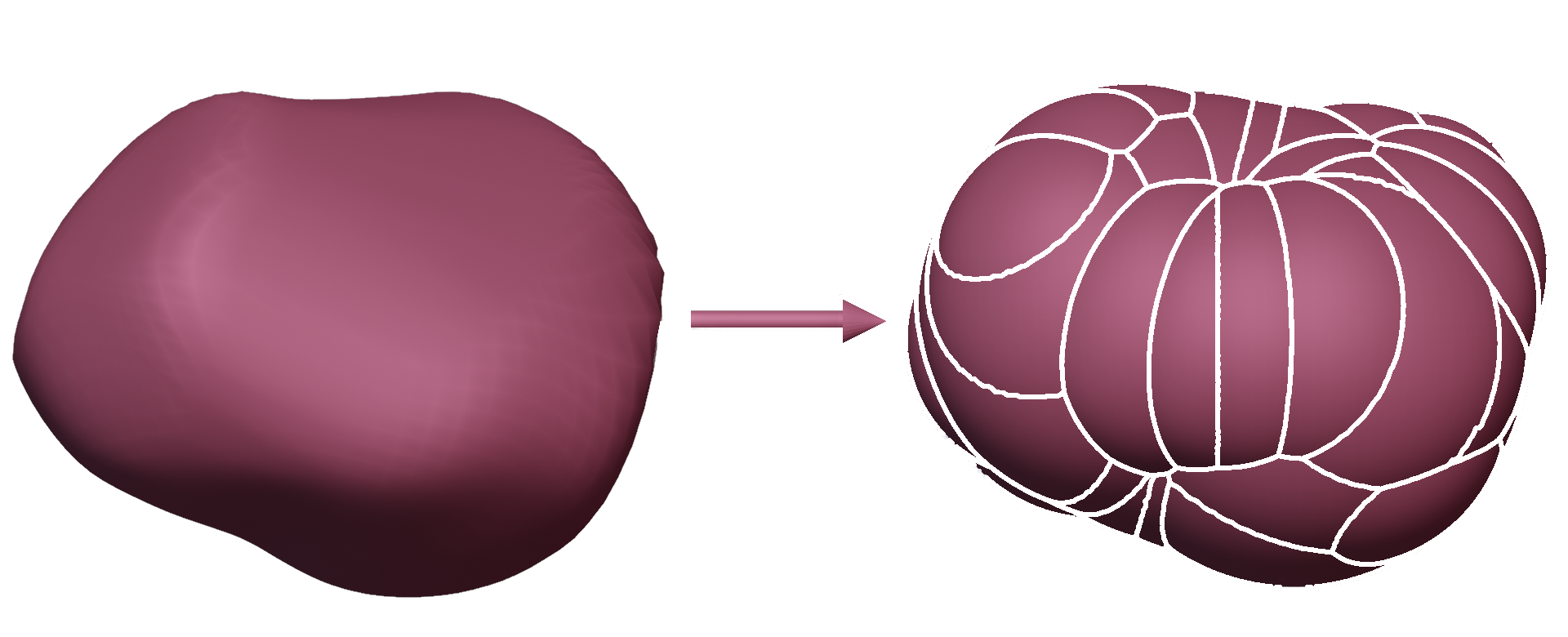}
\caption{A multisphere geometry generated for an arbitrarily shaped non-spherical particle.}
\label{Fig:Multi Sphere Representation}
\end{figure}

Discrete Element Method~(DEM), is a numerical scheme employed to simulate the behavior of particulate/discontinuous material. Its main components include contact detection, force displacement relations, friction, Newton's laws of motion and a time integration scheme to update the position of particles at successive time steps~\cite{cundall1979DEM}.  Spherical particles are most widely simulated using DEM due to the ease of contact detection~\cite{Munjiza1998contactdetection,He2007contact} and the well established contact force-displacement relations~\cite{Hertz1882,Mindlin2021,Johnson1971contact,THORNTON1991153,Alberto2004contact}. To deal with non-spherical particles, two methods are primarily used, superquadric models~\cite{Podlozhnyuk2017superquadric,gao2021superquadric} and multisphere approximation~\cite{li2015multisphere,KRUGGELEMDEN2008153multisphere}. Superquadric particles are a class of particles whose surface can be modelled using the generalised superquadric shape function~(see \cref{sq_equation}~\cite{Podlozhnyuk2017superquadric}). 
\begin{equation}\label{sq_equation}
f(x,y,z)\equiv \left( \abs{\frac{x}{a}}^{n_2} + \abs{\frac{y}{b}}^{n_2} \right) ^{\frac{n_1}{n_2}} + \abs{\frac{z}{c}}^{n_1} - 1 = 0  
\end{equation}
Here, $a,b$ and $c$ are called the scaling parameters as they allow us to scale the dimension of the particles in the three principle axes respectively and $n_1$ and $n_2$ are called the shape parameters which determine the `blockiness' or, inversely the `roundness' of the particles as shown in \cref{Fig: SQ_shapes}. 
Multi Sphere approach is a geometric approximation approach where we replace the complex shape that we wish to simulate by a set of overlapping spheres~(see \cref{Fig:Multi Sphere Representation}). This approximation helps us to use the simple contact detection and force-displacement relations of spherical particles. These two factors allow for easy implementation of multisphere particles in DEM. Hence multisphere approaches are widely used to simulate realistic granular material composed of non-spherical particles like rocks, seeds, pharmaceutical tablets, etc.~\cite{li2015multisphere,markauskas2011investigation,lian2021study,cabiscol2018calibration}.
\begin{figure}[htbp]
\begin{subfigure}[t]{2.866 cm}
\centering\includegraphics[width=.9\textwidth]{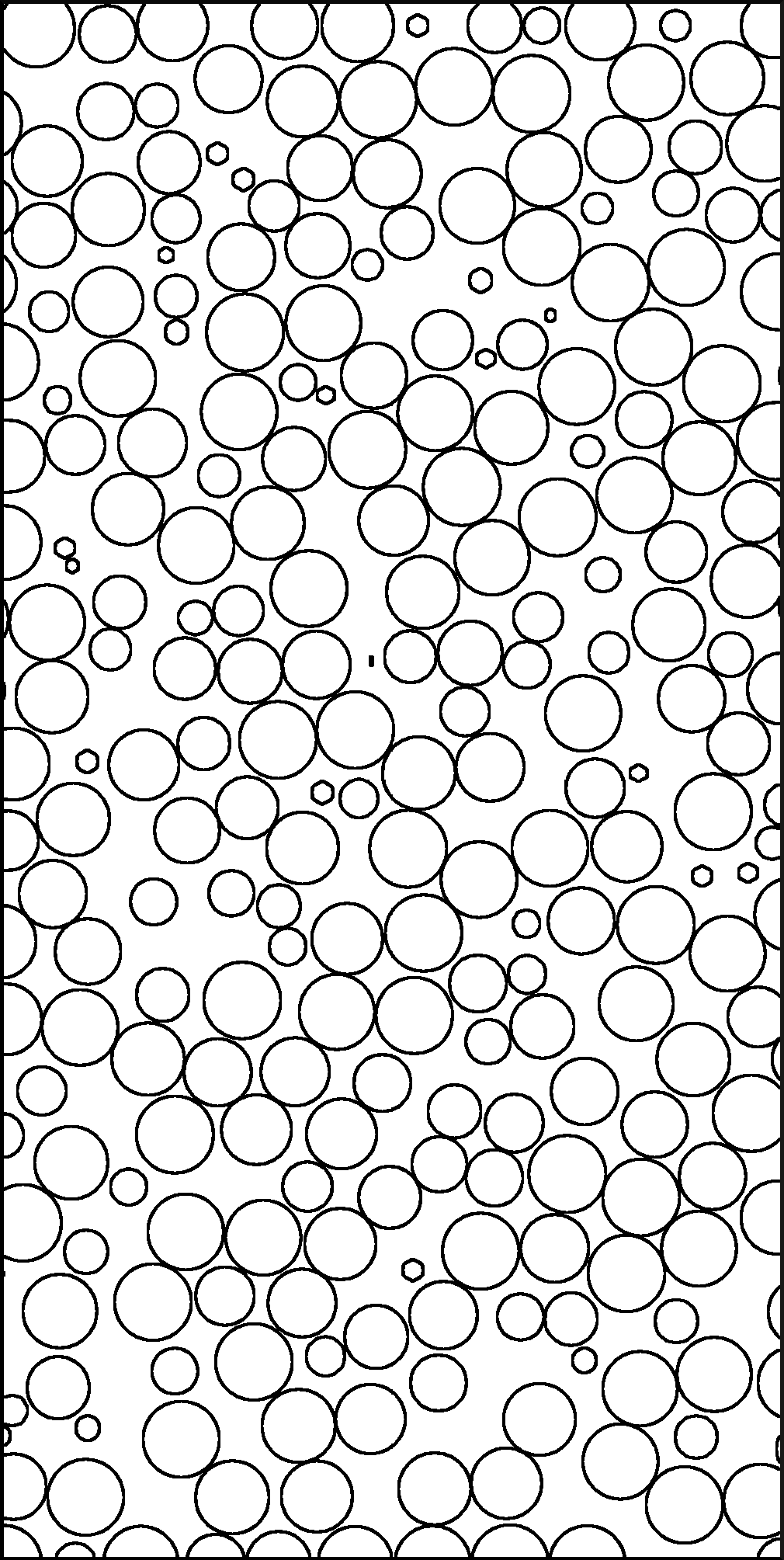}
\subcaption{}
\label{Fig:Slice view circ}
\end{subfigure}%
\begin{subfigure}[t]{2.866 cm}
\centering\includegraphics[width=.9\textwidth]{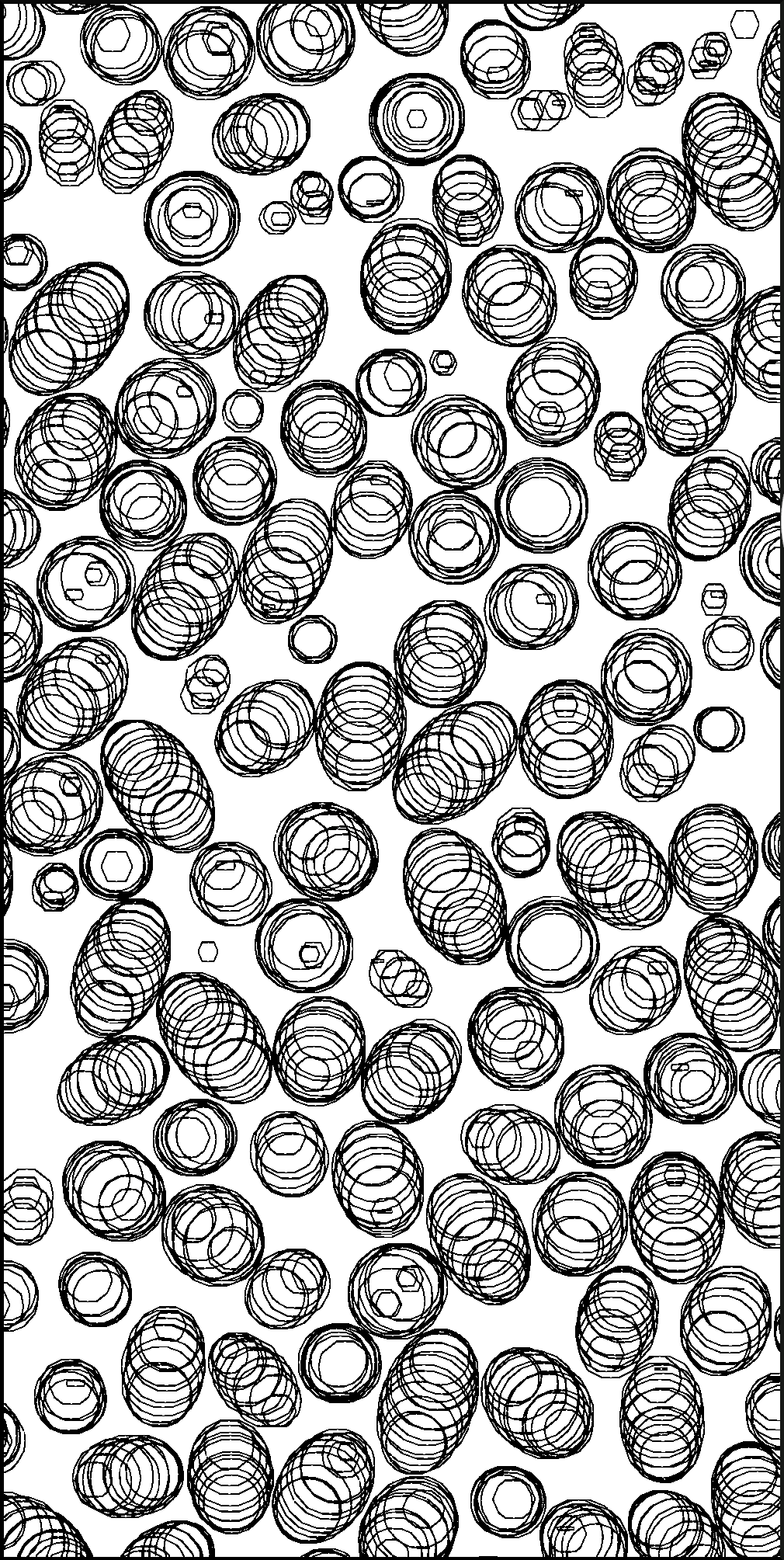}
\subcaption{}
\label{Fig:Slice view elip}
\end{subfigure}%
\begin{subfigure}[t]{2.866 cm}
\centering\includegraphics[width=.9\textwidth]{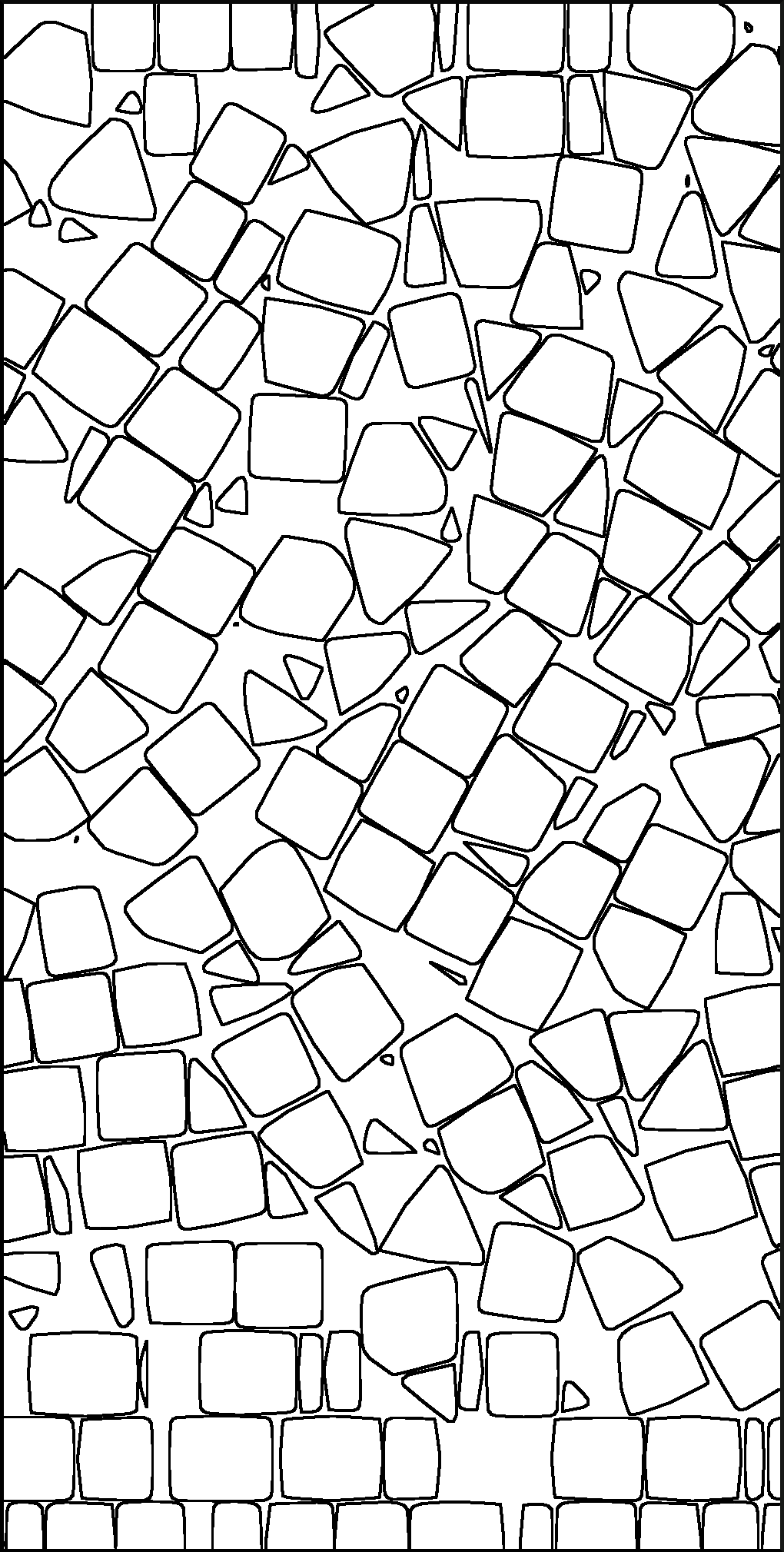}
\subcaption{}
\label{Fig:Slice view super_quadric}
\end{subfigure}
\caption{(a) A sliced view of a particle assembly consisting of spherical particles. (b) The same sliced view for ellipsoidal particles modelled as multi spheres showing the high degree of overlap of spheres.(c) A similar sliced view of superquadric particles with both shape parameters set to a value of 10, the irregular nature of the projected shapes can be clearly observed.}\label{Fig:slice view}
\end{figure}
Studying the packing behaviour of granular assemblies is of prime importance for many industries such as additive manufacturing~\cite{Fouda2019additive,sen2021pharma}, pharmaceuticals~\cite{sen2021pharma,Xiaowei2006pharma}, food grain processing~\cite{Boac2014agri} to study packing efficiency, flow characteristics~\cite{aguirre2014flow}, particle segregation, stability analysis~\cite{olson2005stability}, etc. Factors like presence of walls~\cite{JAGGANNAGARI202131,Kobayakawa2018platedrag}, size distribution of the particles~\cite{Estrada2016grainsize,Mutabaruka2019polydispersity}, shape of the particles~\cite{Olmedilla2019nonconvex}, vibration, various loading conditions~\cite{Kouraytem2016penetration,Tapia2013shearband}, etc. have localised effects in the packing of the particles. The only way to study these effects is by obtaining how the packing density varies across the domain. Studying the  average packing fraction throughout the domain~\cite{CHENG2000123spheresbulk,WANG202129octa,ZHAO2020160tetra,Gan2020cyl} would be useful to study the  behaviour of the particle assembly as a whole, but individual effects of the factors described above cannot be captured and studied from the average packing density.
One  method to characterise the variation of packing of the assembly is to calculate the `variation of packing fraction' of the assembly along any particular direction. This involves calculating the `planar packing fraction' on closely spaced planes along the required direction to get a sense of how the packing varies in that direction. Planar packing fraction is the ratio of the cumulative area of the geometries inscribed onto a plane by the particles to the total area of the plane. There have been a few studies on the packing fraction variation for spherical particles~\cite{gan2010packingpebblebed,Raghuram2018packingprismatic,JAGGANNAGARI202131}. However, such studies on non-spherical particles are not available in the literature to the best of authors' knowledge. It may be because of the challenge posed by the complex shape of the particles, which, in turn, inscribe complicated geometries onto any plane passing through them as shown in \cref{Fig:slice view}.
In this paper, a voxelization based post-processing methodology for DEM simulations is proposed to analyse the packing structure of the assembly. We first create a voxel grid of predefined resolution in the domain and identify a few of those voxels which lie inside the particles so as to accurately capture the geometrical characteristics of the particle assembly~( \cref{sec:Bounding Box Voxelization of Non-Spherical Particles.}). Further we optimise this method of voxelization for multisphere particles. We also introduce an algorithm to compute the `planar packing fraction' along any arbitrary plane in the domain~(\cref{sec:Calculation of Planar Packing Fraction along any Arbitrary Plane}). This gives us the unique ability to analyse the packing structure in any direction.
The voxel data obtained from this algorithm can be used for analysing various characteristics of the assembly. In this paper, we use the voxel data to obtain the packing fraction variation in the three primary directions, as well as generate heat-maps to obtain a scanned view through the assembly when viewed through a particular direction. Also we divide the domain into a few spatial regions and obtain the bulk packing fraction in those regions. Finally, we use our algorithm to calculate the packing fraction in any arbitrary plane to calculate the packing fraction variation along the direction of a face diagonal of the domain and the results are shown in \cref{sec:Packing Fraction Variation for Non-Spherical Particle}
%=================================================%
\section{Bounding box voxelization of non-Spherical particles.}
\label{sec:Bounding Box Voxelization of Non-Spherical Particles.}
\begin{figure}[htbp]
    \centering
    \begin{subfigure}[t]{4.3 cm}
    \centering
    \includegraphics[width=0.8\linewidth]{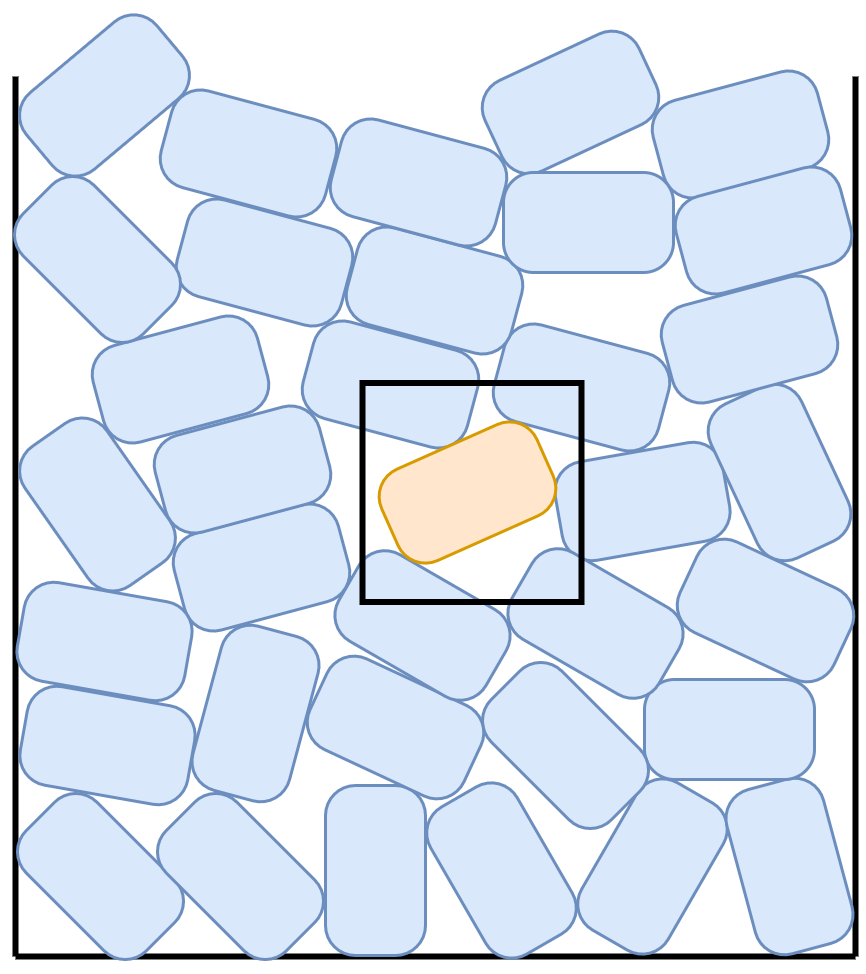}
    \caption{}
    \label{Fig:sqvoxo}
    \end{subfigure}%
    \begin{subfigure}[t]{4.3cm}
    \centering
    \includegraphics[width=0.8\linewidth]{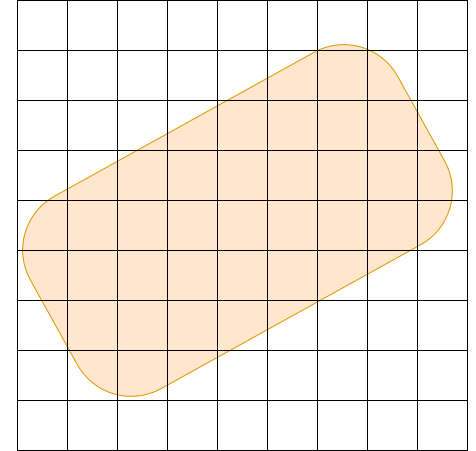}
    \caption{}
    \label{Fig:sqvox1}
    \end{subfigure}
    \begin{subfigure}[t]{4.3 cm}
    \centering
    \includegraphics[width=0.8\linewidth]{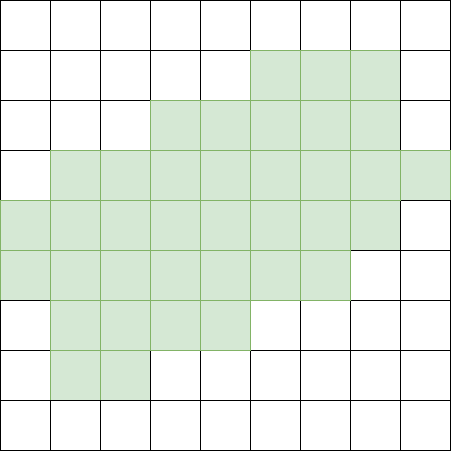}
    \caption{}
    \label{Fig:sqvoxo2}
    \end{subfigure}%
    \begin{subfigure}[t]{4.3 cm}
    \centering
    \includegraphics[width=0.8\linewidth]{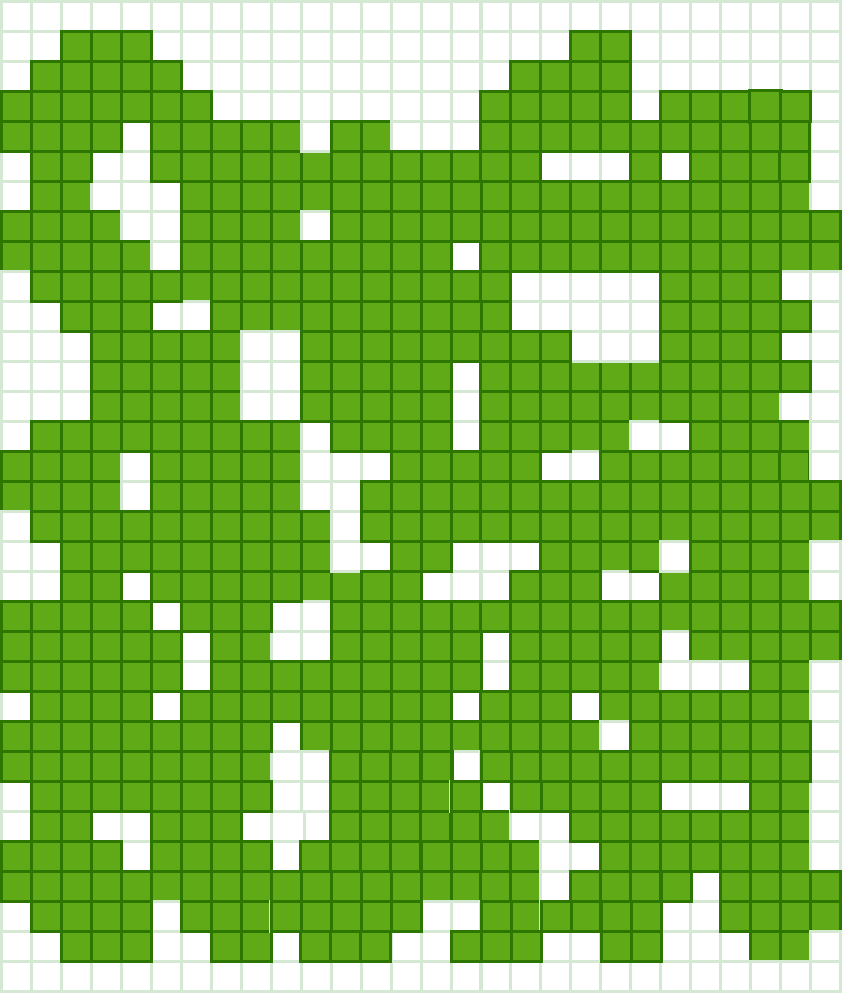}
    \caption{}
    \label{fig:my_label2}
    \end{subfigure}     
    \caption{A simple representation of the steps involved in the voxelization of superquadric particles. (a) The DEM data obtained from the simulation. A single particle along with its bounding box in highlighted. (b) A zoomed in view of the highlighted particle, along with the bounding box containing the voxels. (c) Highlighting the voxels inside the particle. (d) Final voxel data after performing the preceding step for all the particles.}
    \label{Fig:Voxellizastion_sq}
\end{figure}

In this section, we introduce the algorithm for voxelizing the particle data obtained from DEM simulations. The generalised algorithm that is used for both superquadric particles and multisphere particles is presented. Then, we further optimize the voxelization of multisphere particles due the regular nature of the shape of a sphere. The first step of this algorithm is to discretize the domain into cubic volume elements, i.e., voxels. For the discretization, we need the dimensions of the domain and the size of the voxel (here, size corresponds to the edge length of the voxel). We use these values to calculate the number of voxels along the $x,y$ and $z$ directions, denoted by $N_x$, $N_y$ and $N_z$, respectively, resulting in a total of $N_x \times N_y \times N_z$ number of voxels. We represent the `grid' of voxels as a 3D Numpy~\cite{harris2020array} array of shape $[N_x \times N_y \times N_z]$. The second step of the algorithm is identify the voxels which lie inside the particles. To do this, we go through each particle and assign a bounding box to the particle. The bounding box for a particle is a cube centered at the centroid of the particle and has an edge length slightly bigger (5\% in our case) than the largest radial extent of the particle. This ensures that the geometry of the particle does not protrude outside the bounding box, irrespective of the orientation of the particle. After assigning the bounding box, we then isolate the voxels inside the bounding box and check if they are inside or outside the particle. If the voxel is in inside the particle, its value in the voxel array is changed from 0 to 1. An illustration of this procedure is shown in \cref{Fig:Voxellizastion_sq}.
To check if a voxel is inside or outside the particle, We transform the centre of the voxel $(x_{\text{v}},y_{\text{v}},z_{\text{v}})$ into the local coordinate system of the particle, i.e., the coordinate system where the particle can be represented by \cref{sq_equation}. The transformed centre $(x_{\text{v}}^*,y_{\text{v}}^*,z_{\text{v}}^*)$ is plugged into \cref{sq_equation}, which can result in one of the following outcomes.
\begin{enumerate}
    \item $f(x_{\text{v}}^*,y_{\text{v}}^*,z_{\text{v}}^*) < 0$, meaning the centre is inside the particle.
    \item $f(x_{\text{v}}^*,y_{\text{v}}^*,z_{\text{v}}^*) = 0$, meaning the centre is on the surface of the particle.
    \item $f(x_{\text{v}}^*,y_{\text{v}}^*,z_{\text{v}}^*) > 0$, meaning the centre is outside the particle.
\end{enumerate}
In this method, we assume a voxel is inside a particle if the voxel's centroid is inside the particle. This clearly results in some unwanted volume inclusions and exclusions. But, as it will be seen subsequently, if the voxel grid is fine enough, the accumulated volume error would converge to zero.
\subsection{Optimization of voxelization algorithm for multisphere particles}
\begin{figure}[htbp]
    \centering
    \begin{subfigure}[t]{2.87 cm}
    \centering
    \includegraphics[width=0.95\linewidth]{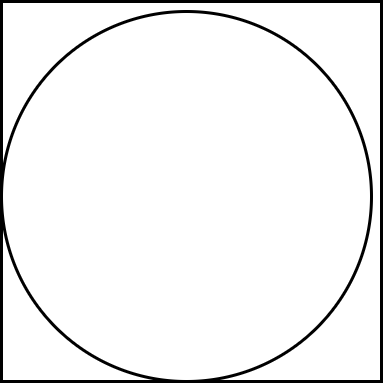}
    \caption{}
    \label{fig: ms_vox1}
    \end{subfigure}%
    \begin{subfigure}[t]{2.87 cm}
    \centering
    \includegraphics[width=0.95\linewidth]{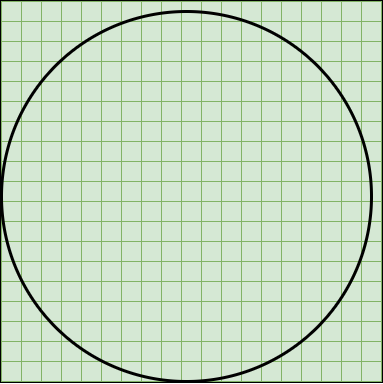}
    \caption{}
    \label{fig: ms_vox2}
    \end{subfigure}%
    \begin{subfigure}[t]{2.87 cm}
    \centering
    \includegraphics[width=0.95\linewidth]{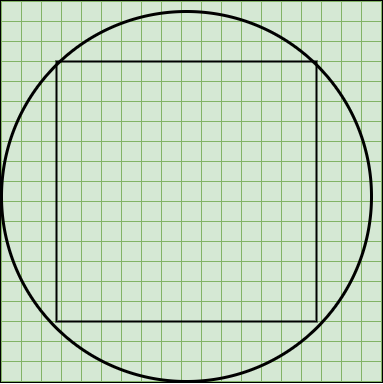}
    \caption{}
    \label{fig: ms_vox3}
    \end{subfigure}
    \begin{subfigure}[t]{2.87 cm}
    \centering
    \includegraphics[width=0.95\linewidth]{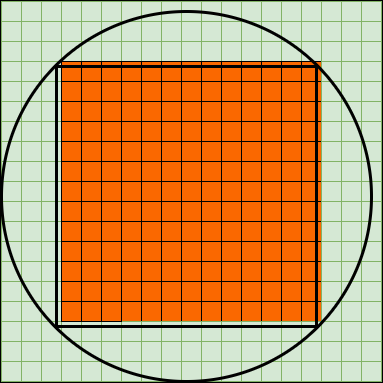}
    \caption{}
    \label{fig: ms_vox4}
    \end{subfigure}%
    \begin{subfigure}[t]{2.87 cm}
    \centering
    \includegraphics[width=0.95\linewidth]{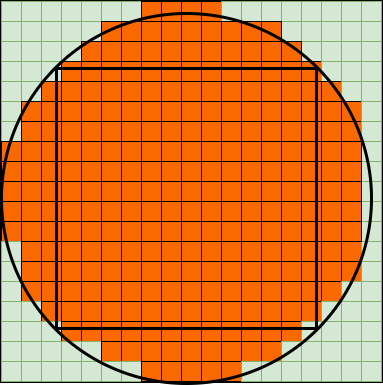}
    \caption{}
    \label{fig: ms_vox5}
    \end{subfigure}%
    \begin{subfigure}[t]{2.87 cm}
    \centering
    \includegraphics[width=0.95\linewidth]{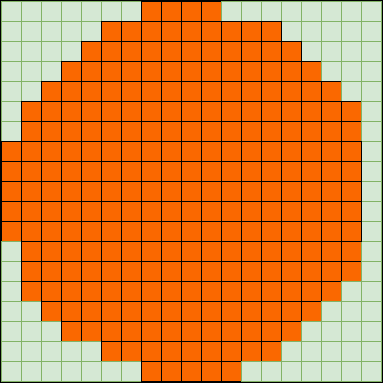}
    \caption{}
    \label{fig: ms_vox6}
    \end{subfigure} 
    \caption{Two-dimensional schematic of the sequence of steps involved in voxelization of a sphere. (a) The sphere along with its bounding box. (b) The voxels inside the bounding box. (c) Isolating the largest cube that can fit inside the sphere. (d) Automatically marking the voxels inside the cube. (e) Checking and marking the rest of the voxels inside the bounding box. (d) The final voxel arrangement.}
    \label{Fig: ms_vox}
\end{figure}

The bounding box method for multisphere particles can be a redundant approach due to the large number of spheres per particle. Depending on the complexity of the shape we are modelling, each particle could be composed of a few tens of spheres to a few hundreds of spheres. Creating a bounding box for each of the sphere and evaluating each voxel inside these boxes would be computationally ineffective. We can counter this by leveraging the regular nature of the shape of a sphere. We first find the largest cube that can fit inside a sphere and fill all the voxels automatically. This can be done because orientation does not effect the spatial occupancy of a sphere, hence the largest cube that can fit inside the particle is not affected by the particle orientation. This leads to filling of nearly 70\% of the voxels automatically, greatly reducing the number of computations required. A graphic representation of the voxelization process is shown in \cref{Fig: ms_vox}. All the codes developed for the algorithm and the results presented in this article are provided for the benefit of the readers at \url{https://github.com/Manojpvr/VoxelDEM} \cite{Pola_VoxelDEM_2021}.
%=================================================%
\section{Validation of voxelization algorithms}
\label{sec:Validation of Voxelization Algorithms}
\begin{figure}[htbp]
    \begin{subfigure}[t]{4.3 cm}
    \centering
    \includegraphics[width=0.7\linewidth]{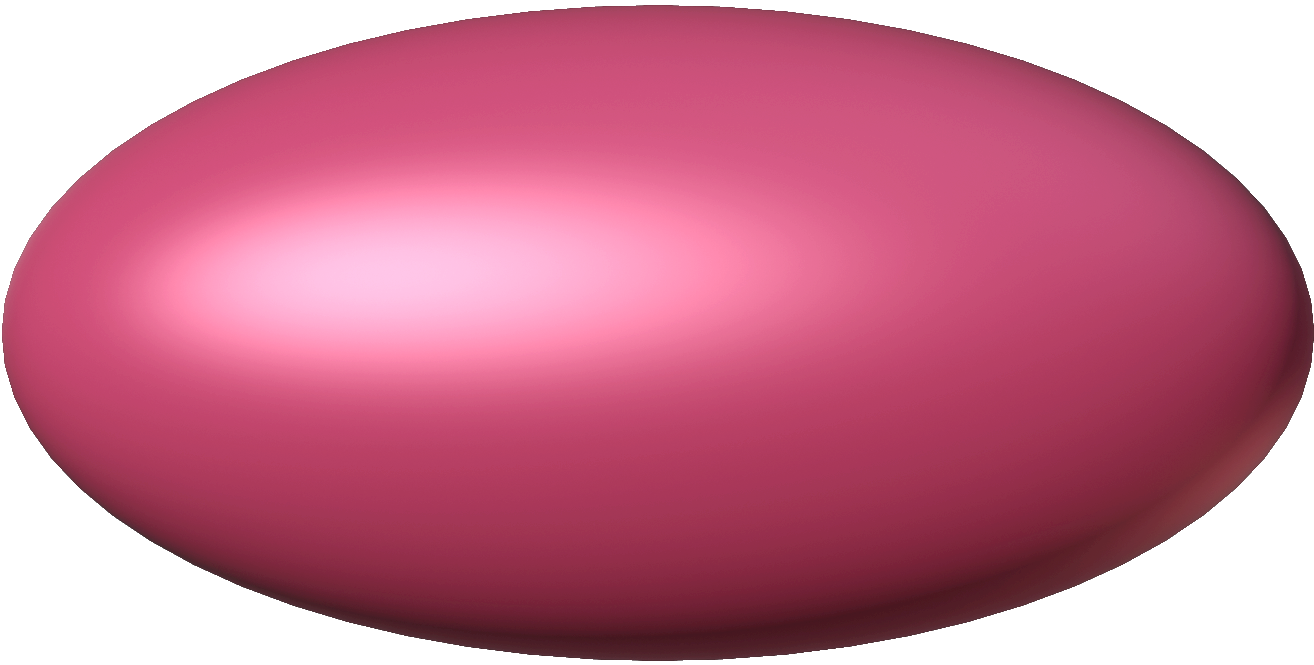}
    \caption{}
    \label{fig: elp_sq}
    \end{subfigure}%
    \centering
    \begin{subfigure}[t]{4.3 cm}
    \centering
    \includegraphics[width=0.7\linewidth]{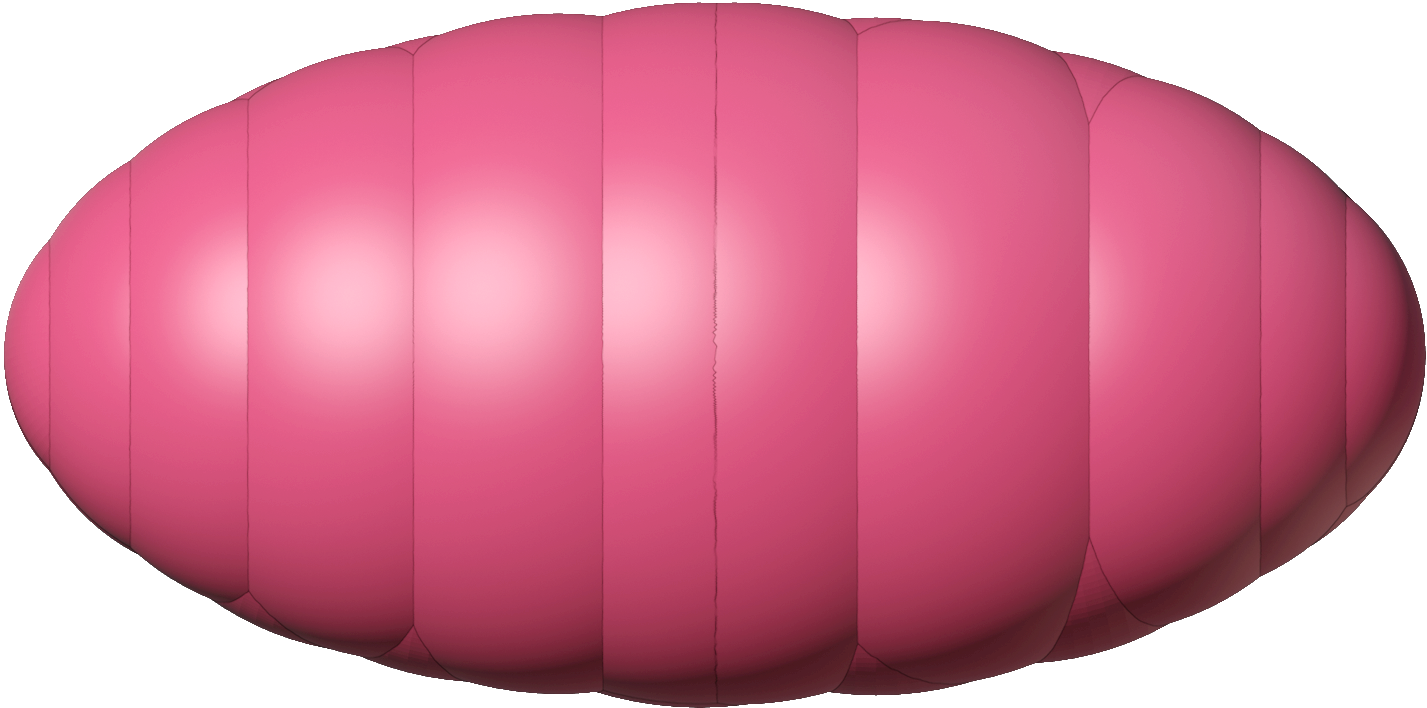}
    \caption{}
    \label{fig: elp_ms}
    \end{subfigure}%
    \caption{(a) Superquadric ellipsoidal particle. (b) multisphere ellipsoidal particle consisting of 20 spheres. These two particle shapes were used for validation and computation time analysis. The striations on the surface of the multi sphere particle represent the interpenetrating spheres.}
    \label{Fig:ms_vs_sq_elp}
\end{figure}

\begin{figure}[htbp]
    \centering
    \begin{subfigure}[t]{8.6 cm}
    \centering
    \includegraphics[width=0.99\linewidth]{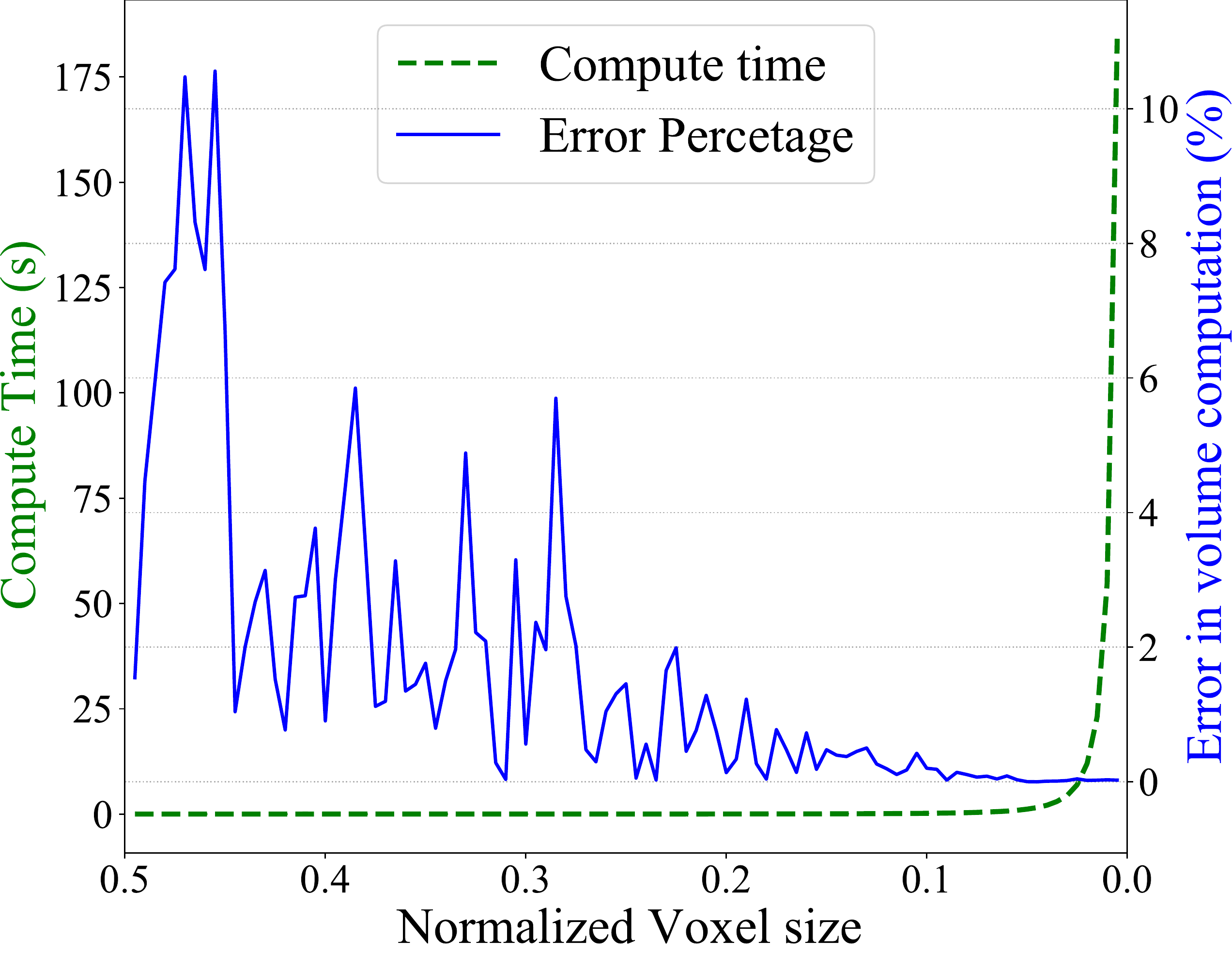}
    \caption{}
    \label{fig: tv_sq}
    \end{subfigure}
    \begin{subfigure}[t]{8.6 cm}
    \centering
    \includegraphics[width=0.99\linewidth]{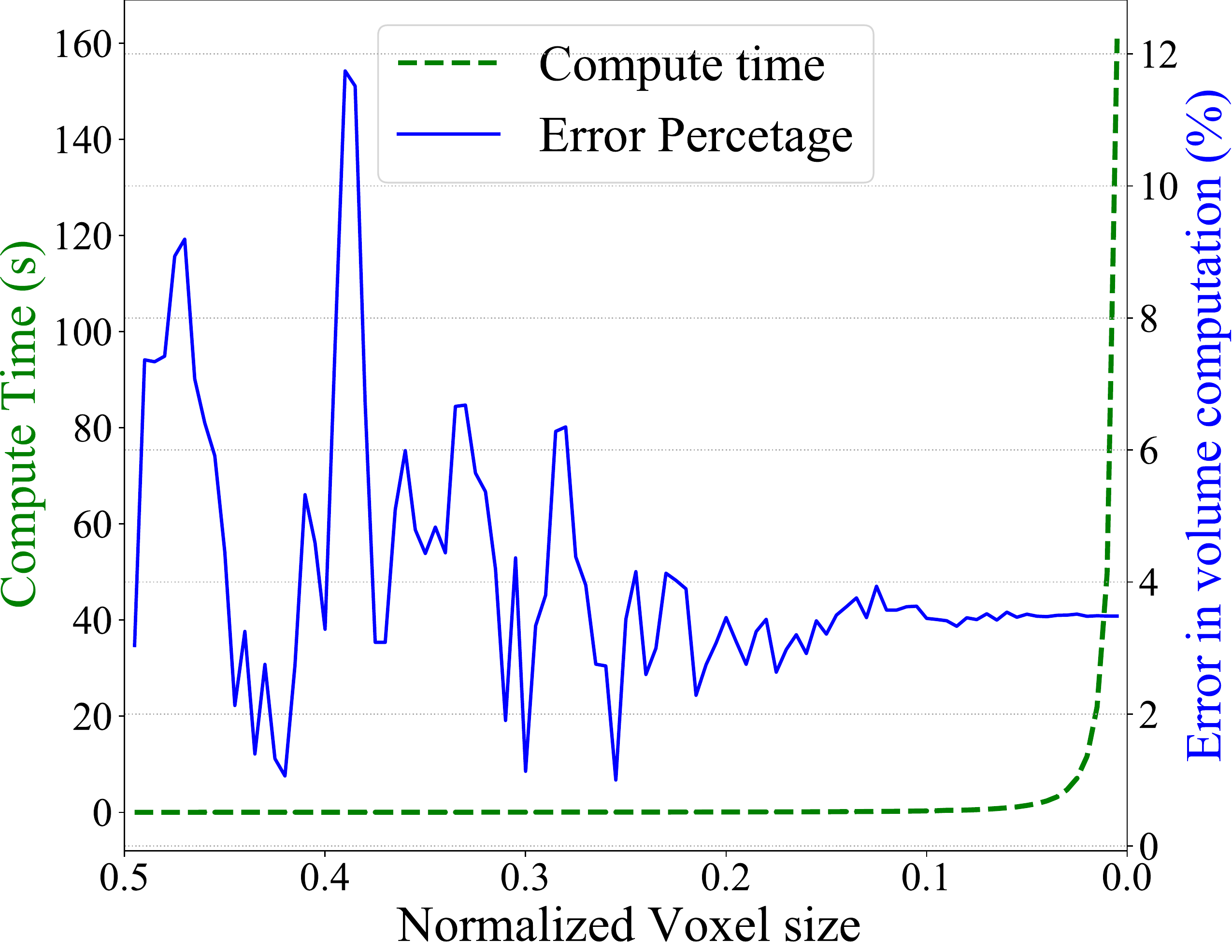}
    \caption{}
    \label{fig: tv_ms}
    \end{subfigure}%
    \caption{The trade-off between computation time and geometric accuracy in voxelizing a prolate spheroid of semi-minor axis $\SI{0.6}{\milli\meter}$ and semi-major axis $\SI{1.12}{\milli\meter}$ when the particle is modelled as (a) superquadric particle (b) multisphere particle consisting of 20 spheres.}
    \label{Fig:time_validation}
\end{figure}

In this section, we validate our algorithm by using it to compute the volume of an ellipsoidal particle. We voxelize the particle and compute the cumulative volume of the voxels inside the particle, and compare the volume with the analytical volume of the ellipsoid. We have modelled the ellipsoidal particle both as a superquadric and as a multisphere as shown in \cref{Fig:ms_vs_sq_elp}.
The normalised voxel size, quantified as the ratio of the edge length of the voxel to minor semi-axis of the particle, is varied from 0.5 to 0.05. The percentage error in calculating the volume of the particle from the voxel data and the computation time for both the cases is shown in \cref{Fig:time_validation}.
From \cref{Fig:time_validation}, we can see uniform convergence for normalised voxel size less than 0.1. For all the simulations presented in this paper, we set the normalised voxel size of 0.1 as it is a good trade-off between accuracy and computation time. One observation from the error convergence plots is that, the volume error for the superquadric particle is converging to 0, whereas, for multisphere particle, it is converging close to 4\%. The convergence error in the multisphere approach is due to the fact that the geometry of the ellipsoid formed by multisphere approach would slightly deviate from a true ellipsoid (see \cref{Fig:ms_vs_sq_elp}) even if a very large number of spheres are used to build the multisphere.
%=================================================%

\section{Calculation of planar packing fraction along any arbitrary plane}
\label{sec:Calculation of Planar Packing Fraction along any Arbitrary Plane}
\begin{figure}[htb]
\begin{center}
\includegraphics[width= 8.6 cm]{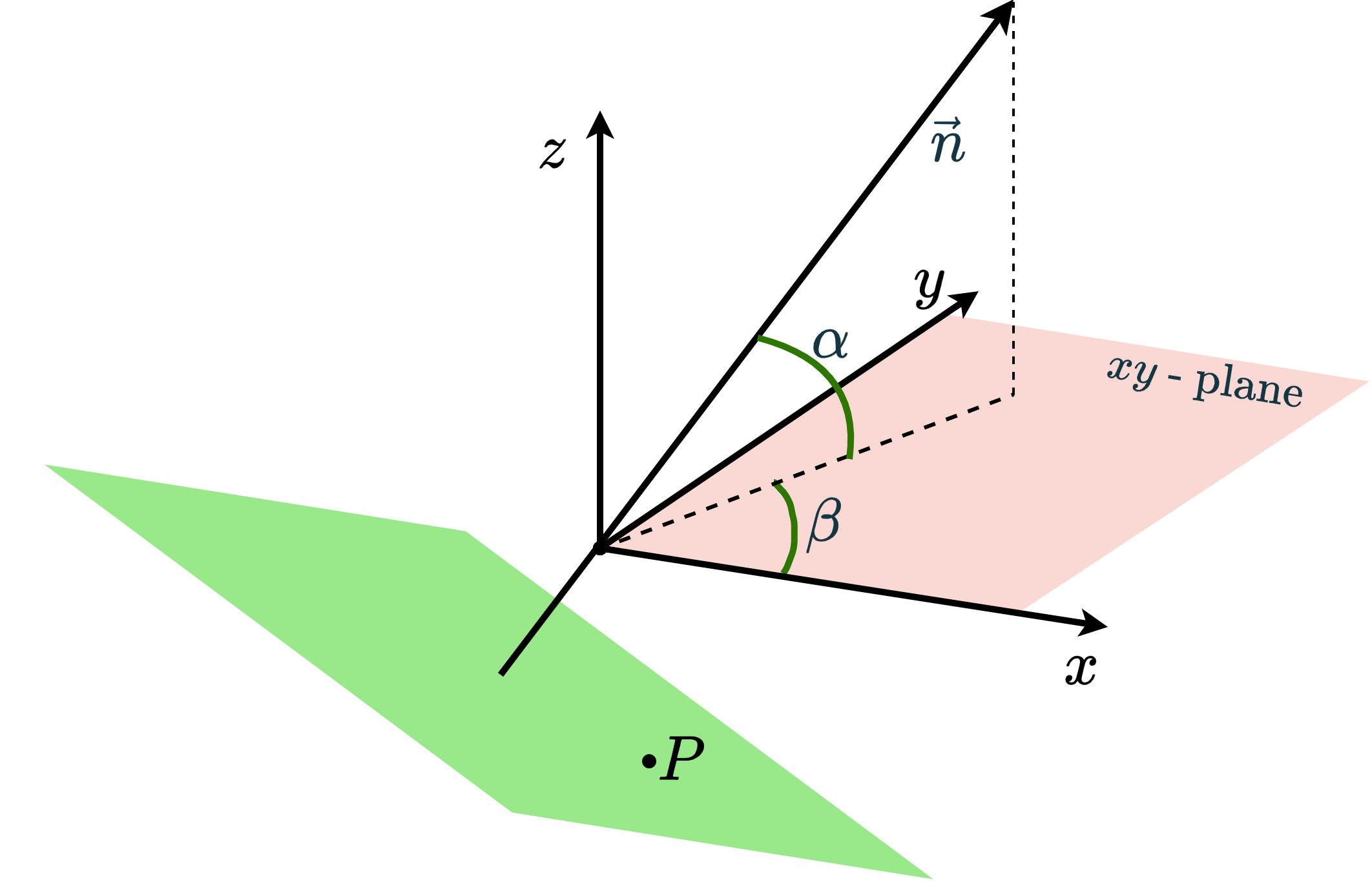}
\caption{The point-normal visualisation of the plane (coloured green) on which the packing fraction is to be estimated.}
\label{Fig:Plane}
\end{center}
\end{figure}

\begin{figure}[htbp]
    \begin{subfigure}[t]{4.3 cm}
    \centering
    \includegraphics[width=0.95\linewidth]{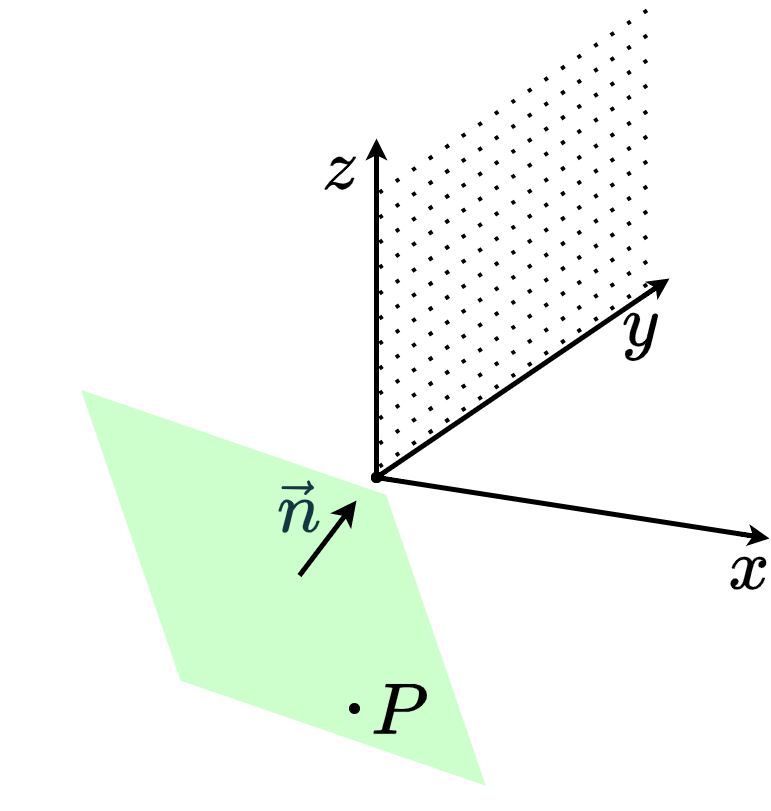}
    \caption{}
    \label{fig: plane_transfom_1}
    \end{subfigure}%
    \centering
    \begin{subfigure}[t]{4.3 cm}
    \centering
    \includegraphics[width=0.95\linewidth]{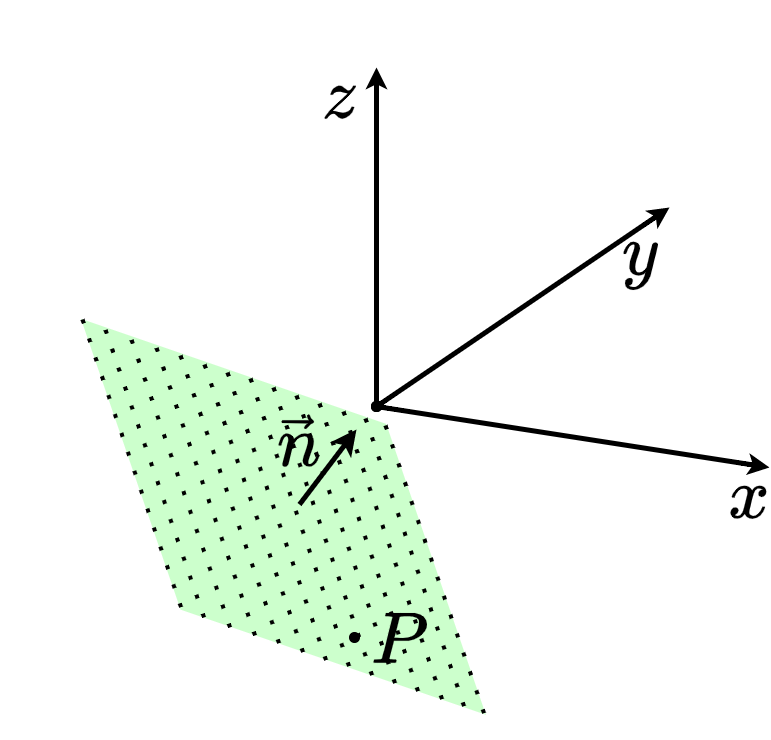}
    \caption{}
    \label{fig: plane_transform_2}
    \end{subfigure}%
    \caption{Transformation of the points from the yz-plane to the desired plane. (a) The original grid of points on the yz-plane(b) The final grid of points on the desired plane.}
    \label{Fig:plane transform}
\end{figure}

Planar packing fraction is the measure of the `packing density' of an assembly of particles on a particular plane. It is calculated as the ratio of the cumulative area of the geometric intersections of the particles in the plane to the total area of the plane. From the voxel data, computing the packing fraction in a plane oriented along either the x, y or z direction is trivial, as these are the directions of voxel index progression. Numpy allows us to slice the voxel array in a way as to easily isolate the voxels through which these planes pass through. In this section, we present an algorithm to compute the packing fraction along an arbitrary plane from the voxel data that has been generated by the preceding algorithm.
The first step to calculate the planar packing fraction on a selected arbitrary plane is to identify a grid of points on the plane. One way to do this would be to first generate points on any known plane(we have chosen the yz plane here), and then transform the points onto the required arbitrary plane. For example, consider a plane described by its normal $\vec{n}$ and passing  through the point $P$ as shown in \cref{Fig:Plane}. Here, the normal vector $\vec{n}$ makes and angle $\alpha$ with its projection on the xy-plane and its projection on the xy-plane makes and angle $\beta$ with the x-axis. Since the points are on the yz-plane, whose normal is the x-axis, one possible transformation that will move the points onto the concerned plane is the rotation which will make the x-axis parallel to $\vec{n}$, which can be seen in \cref{eq:combined} and the translation which will shift the resultant yz-plane to coincide with point $P$ (\cref{eq:point_trans}).
\begin{equation}\label{eq:combined}
     T = \begin{bmatrix}
\cos\beta & -\sin\beta & 0 \\
\sin\beta & \cos\beta & 0 \\
0 & 0 & 1\\
\end{bmatrix} \begin{bmatrix}
\cos(-\alpha) & 0  & \sin(-\alpha) \\
0 & 1 & 0\\
-\sin(-\alpha) & 0 & \cos(-\alpha) \\
\end{bmatrix}
\end{equation}
\begin{equation}\label{eq:point_trans}
    \vec{X}_{\text{final}} = \vec{P} + T\vec{X}_{\text{initial}}
\end{equation}
Here, $\vec{X}_{\text{initial}}$, $\vec{X}_{\text{final}}$ and $\vec{P}$ are position vectors of points prior to the transformation, points obtained after transformation and the point P respectively. After applying this transformation to all the points on the grid we generated on the yz-plane, we get the transformed grid on the plane of consideration as shown in~\cref{Fig:plane transform}.
The next step is to remove all the points outside the domain. It is to be noted that the initial grid should be large enough to slice the domain completely irrespective of the transformation applied. Once we have the leftover points, we need to check the voxel in which each point lies in, and the number of those voxels which are inside the particles. This would give us all the voxels which intersect the plane. The packing fraction on the plane would be the ratio of the number of voxels in the plane which lie inside the particles to the total number of voxels in the plane. For this algorithm to work accurately, it is important that the initial points generated on the $yz$-plane are denser than the voxel grid, i.e, the distance between any two neighbouring points should be less than the edge length of the voxel. Using the procedure described as above, we can now compute the planar packing fraction in any plane that slices the domain with particles of any random shape.
%
%=================================================%
\section{Packing analysis of non-spherical particle assemblies}
\label{sec:Packing Fraction Variation for Non-Spherical Particle}
\begin{figure}[htbp]
    \centering
    \begin{subfigure}[t]{2.87 cm}
    \centering
    \includegraphics[width=0.7\linewidth]{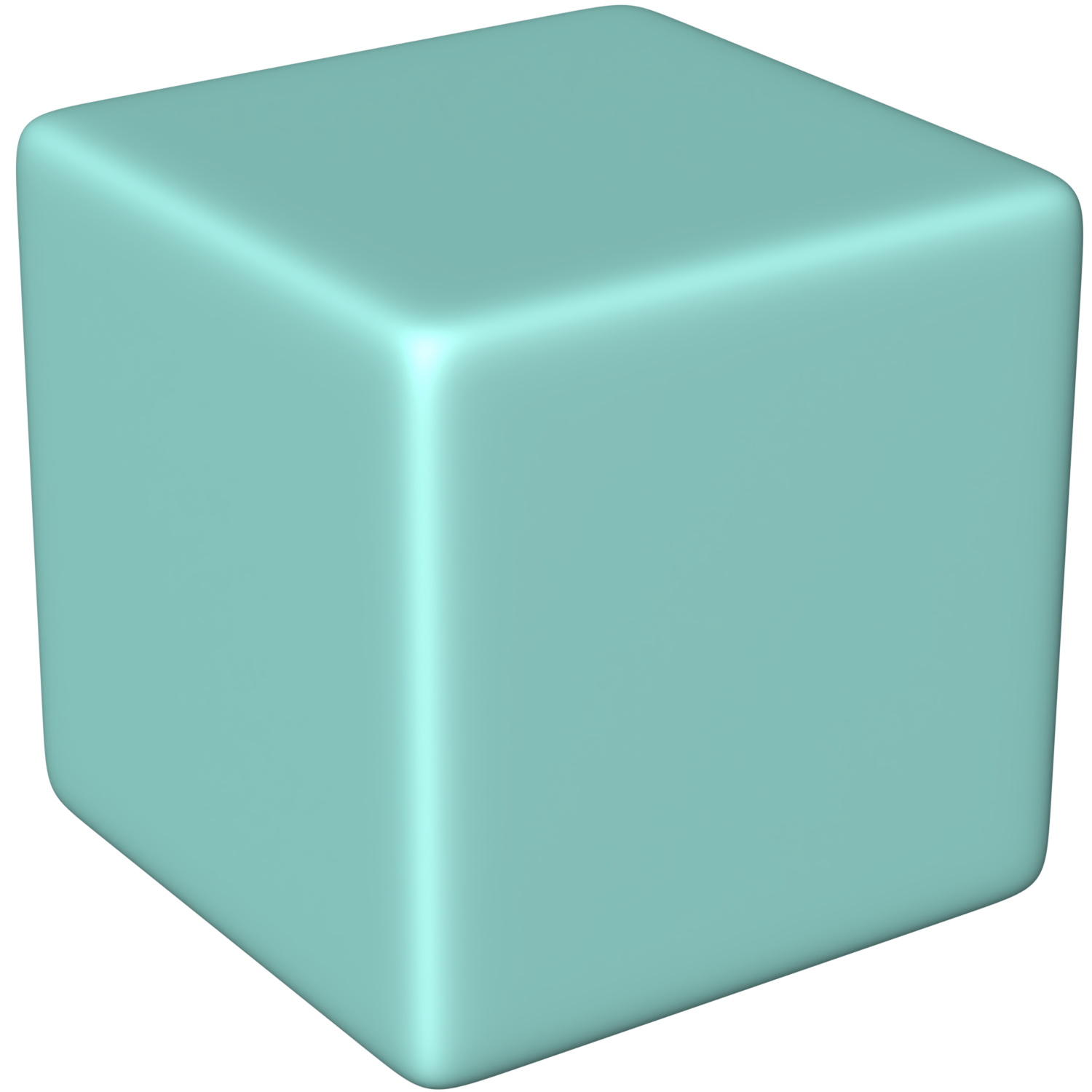}
    \caption{}
    \label{fig: par case 1}
    \end{subfigure}%
    \begin{subfigure}[t]{2.87 cm}
    \centering
    \includegraphics[width=0.7\linewidth]{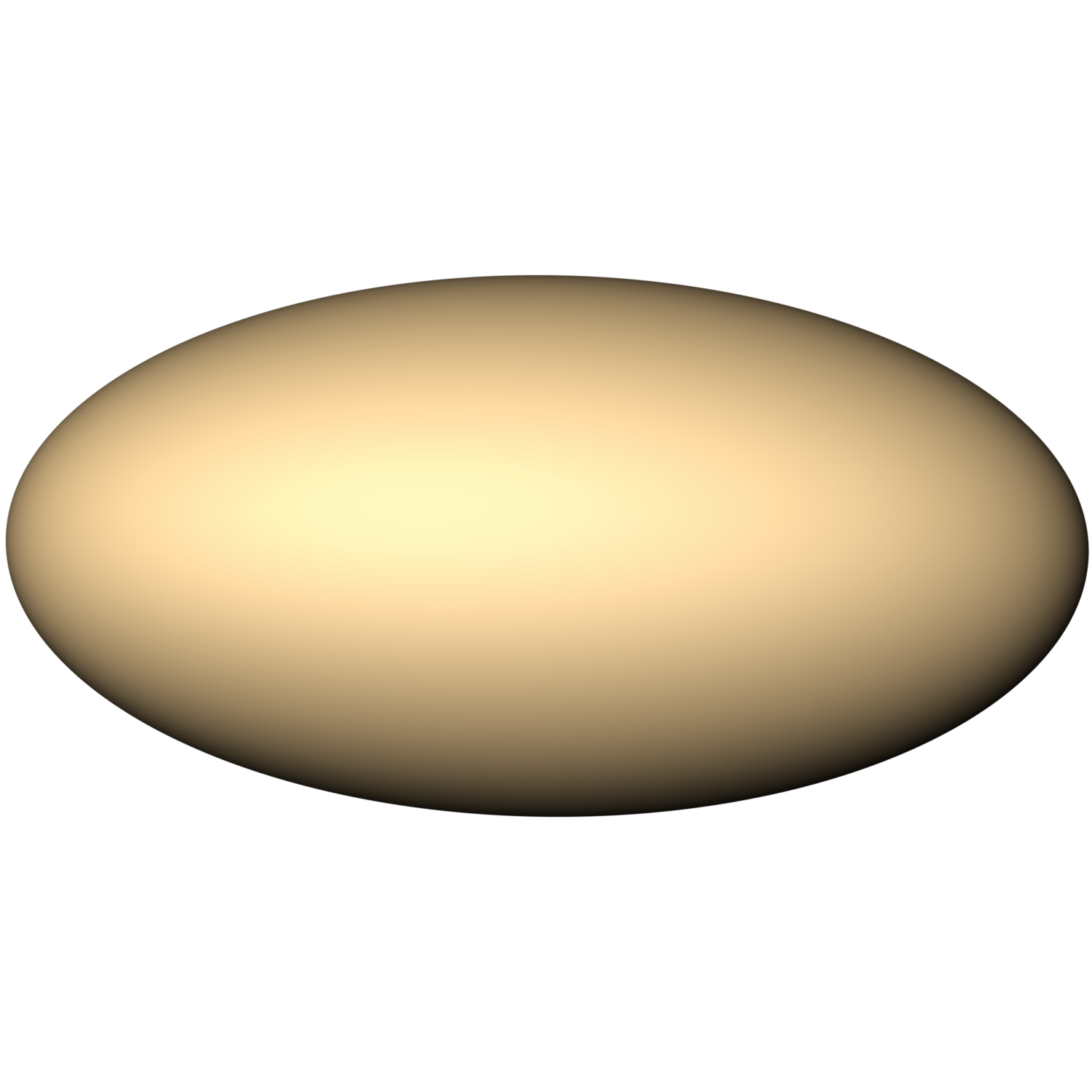}
    \caption{}
    \label{fig: par case 2}
    \end{subfigure}%
    \begin{subfigure}[t]{2.87 cm}
    \centering
    \includegraphics[width=0.7\linewidth]{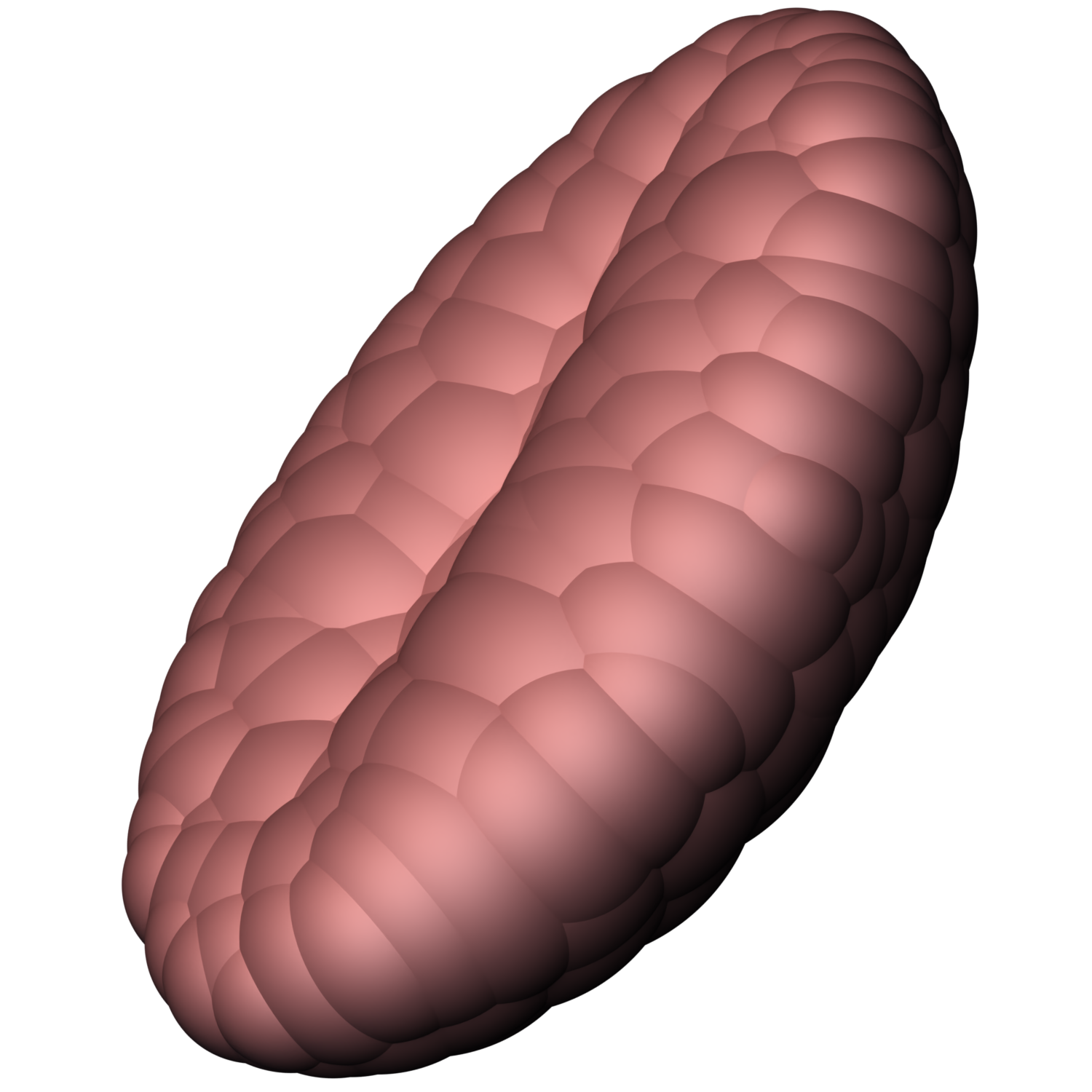}
    \caption{}
    \label{fig: par case 3}
    \end{subfigure}
    \centering
    \begin{subfigure}[t]{2.87 cm}
    \centering
    \includegraphics[width=0.99\linewidth]{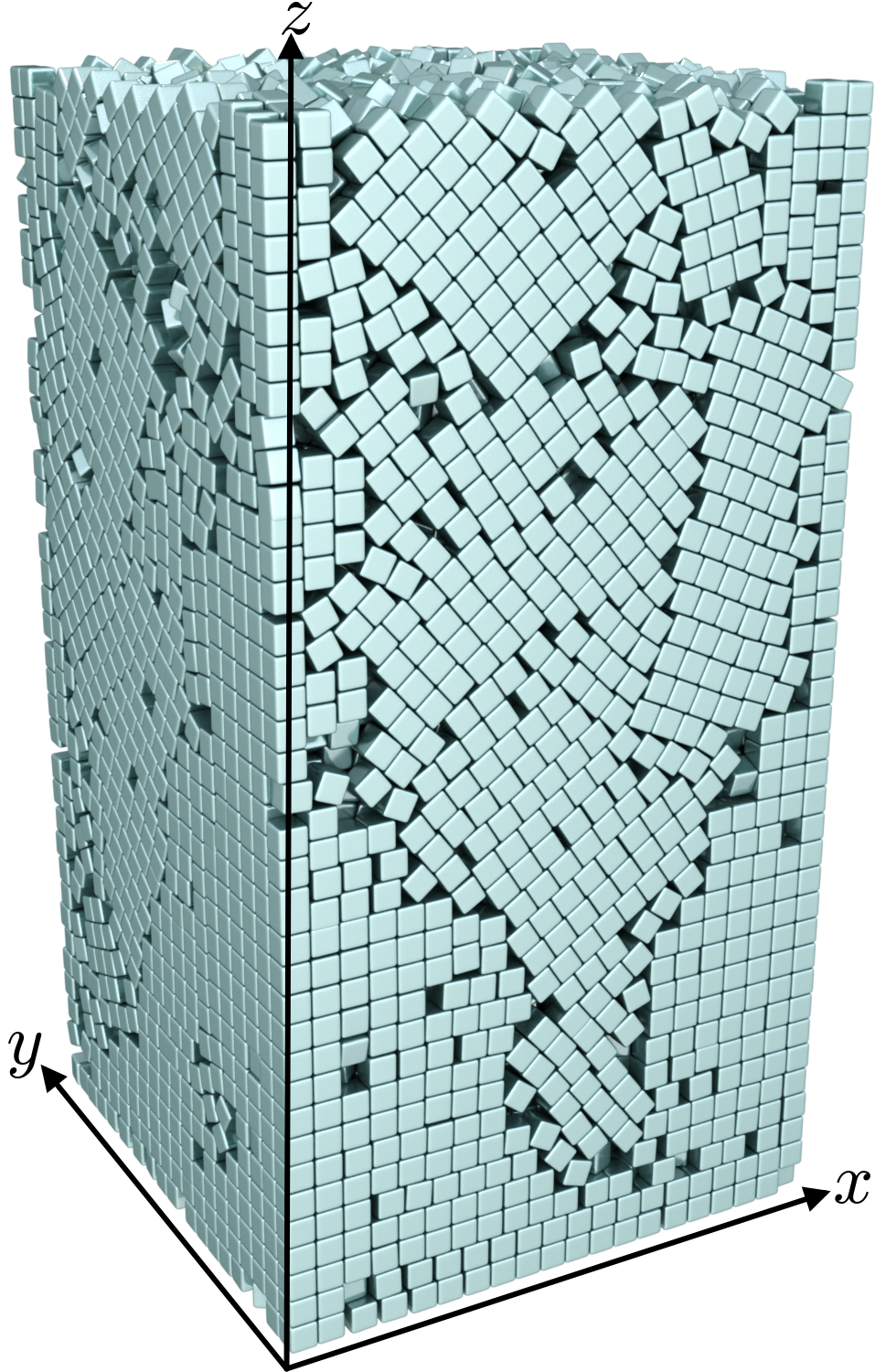}
    \caption{}
    \label{fig: par case 1 as}
    \end{subfigure}%
    \begin{subfigure}[t]{2.87 cm}
    \centering
    \includegraphics[width=0.99\linewidth]{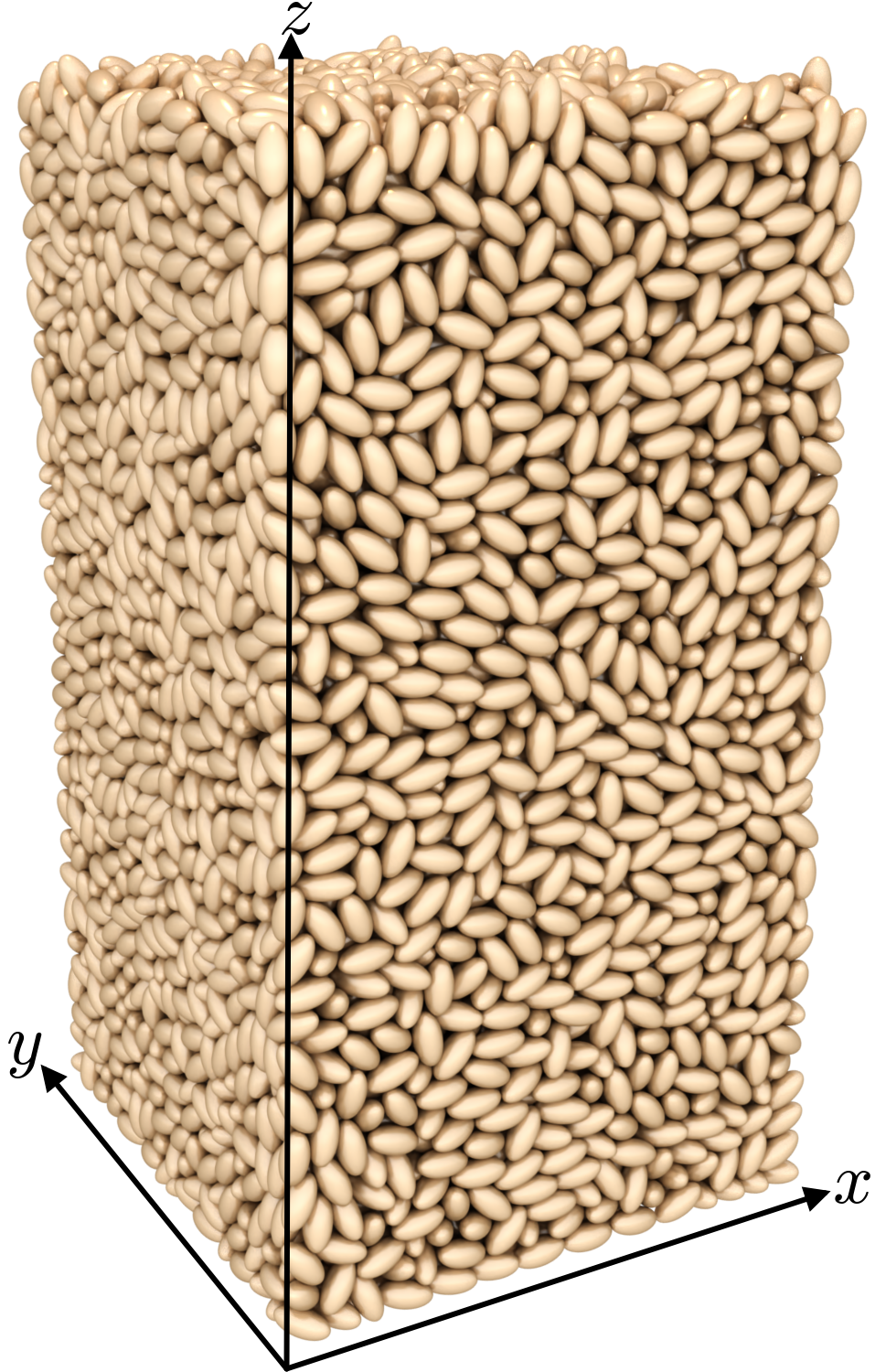}
    \caption{}
    \label{fig: par case 2 as}
    \end{subfigure}%
    \begin{subfigure}[t]{2.87 cm}
    \centering
    \includegraphics[width=0.99\linewidth]{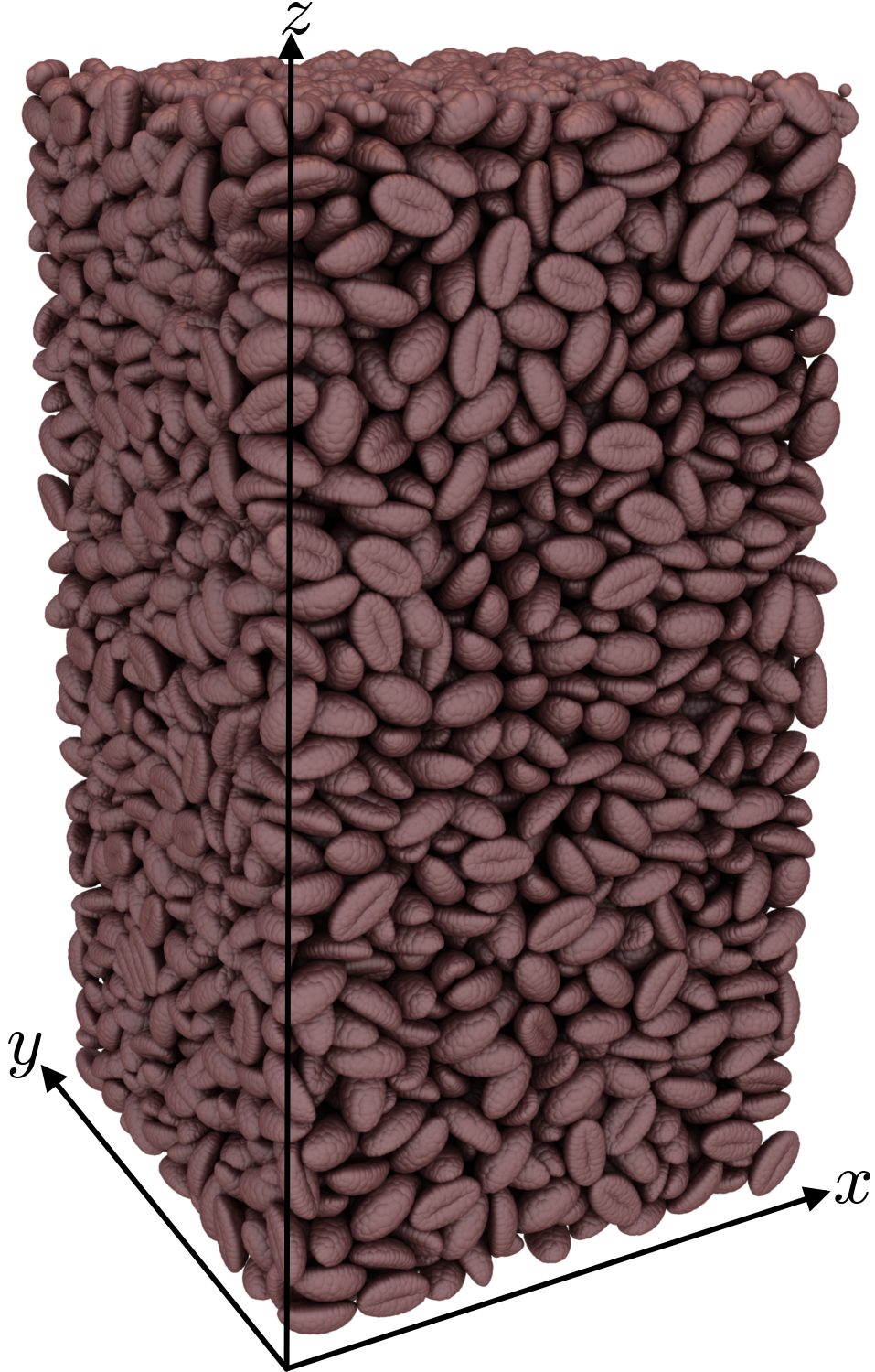}
    \caption{}
    \label{fig: par case 3 as}
    \end{subfigure}%
    \caption{The three particle shapes along with the respective granular assembly used to demonstrate the voxelization algorithm, (a) Cubic superquadric particle (b) Ellipsoidal superquadric particle (c) Coffee bean shaped multisphere particle (d) Cubic superquadric particle assembly (e) Ellipsoidal superquadric particle assembly (f) Coffee bean shaped multisphere particle assembly.}
    \label{fig: RES PAR}
\end{figure}

In this section, we use the algorithms described above to analyse the packing behaviour of granular assemblies (in a cuboid container) of various shapes modelled as both superquadrics and multi spheres. The first step is to the run the DEM simulations to obtain the particle assembly for the desired post processing. We have simulated a total of three cases, two cases of superquadric particles and one case of multisphere particles.
\begin{itemize}
    \item Case 1: Superquadric particle with $n_1=n_2=10$ and $a=b=c=\SI{0.8}{\milli\meter}$, to model nearly cubic particles with an edge length of \SI{1.6}{mm} (see \cref{fig: par case 1} and \cref{fig: par case 1 as}).

    \item Case 2: Superquadric Particles with $n_1 = n_2 = 2$ and $a = b = \SI{0.6}{\milli\meter}$ and $c = \SI{1.2}{\milli\meter}$ to model prolate spheroidal particles with semi-minor axis equal to $\SI{0.6}{mm}$ and semi-major axis equal to $\SI{1.2}{mm}$ (see \cref{fig: par case 2} and \cref{fig: par case 2 as}).
    
    \item Case 3: A coffee bean shaped particle constructed using the multisphere approach (see \cref{fig: par case 3} and \cref{fig: par case 3 as}).

\end{itemize}
In each case, we have simulated pouring of the particles under gravity into a square prismatic container of size $\SI{25}{mm}\times\SI{25}{mm}\times\SI{45}{mm}$. The particles are filled to a height above the marked height to overcome the heaping effect. The material properties used for the particles can be seen in \cref{Table: Material Properties}. The pouring of the multisphere particles was simulated using the Altair EDEM\textsuperscript{\textregistered} program, where as the superquadric particle pouring was simulated using the open source discrete element method program LIGGGHTS~\cite{LIGGGHTS}.
\begin{table}[h]
\caption{Material Properties}
\centering
\begin{tabular}{L{5cm} L{2cm} }
\hline
{\bfseries Property} & {\bfseries Value}\\
\hline
Young's modulus & 
\SI{70}{\giga\pascal} \\
Poisson's ratio & 0.22\\
Density & \SI[per-mode=symbol]{2500}{\kilo\gram\per\meter\cubed}\\
Coefficient of restitution    & 0.5\\
Coefficient of friction & 0.5\\
Coefficient of rolling friction & 0.01\\
\hline
\end{tabular}
\label{Table: Material Properties}
\end{table}
For each case, we voxelized the assembly with a normalized voxel size of the 0.1 (for the superquadrics, the voxel size was normalised with the smallest scaling factor of the particle and for the coffee bean shaped particle, the voxel size was normalised with half the width of the bean). After obtaining the voxel data, we perform the following types of packing fraction computations and present the results.

\begin{figure}[htbp]
\begin{center}
\includegraphics[width= 7.5 cm]{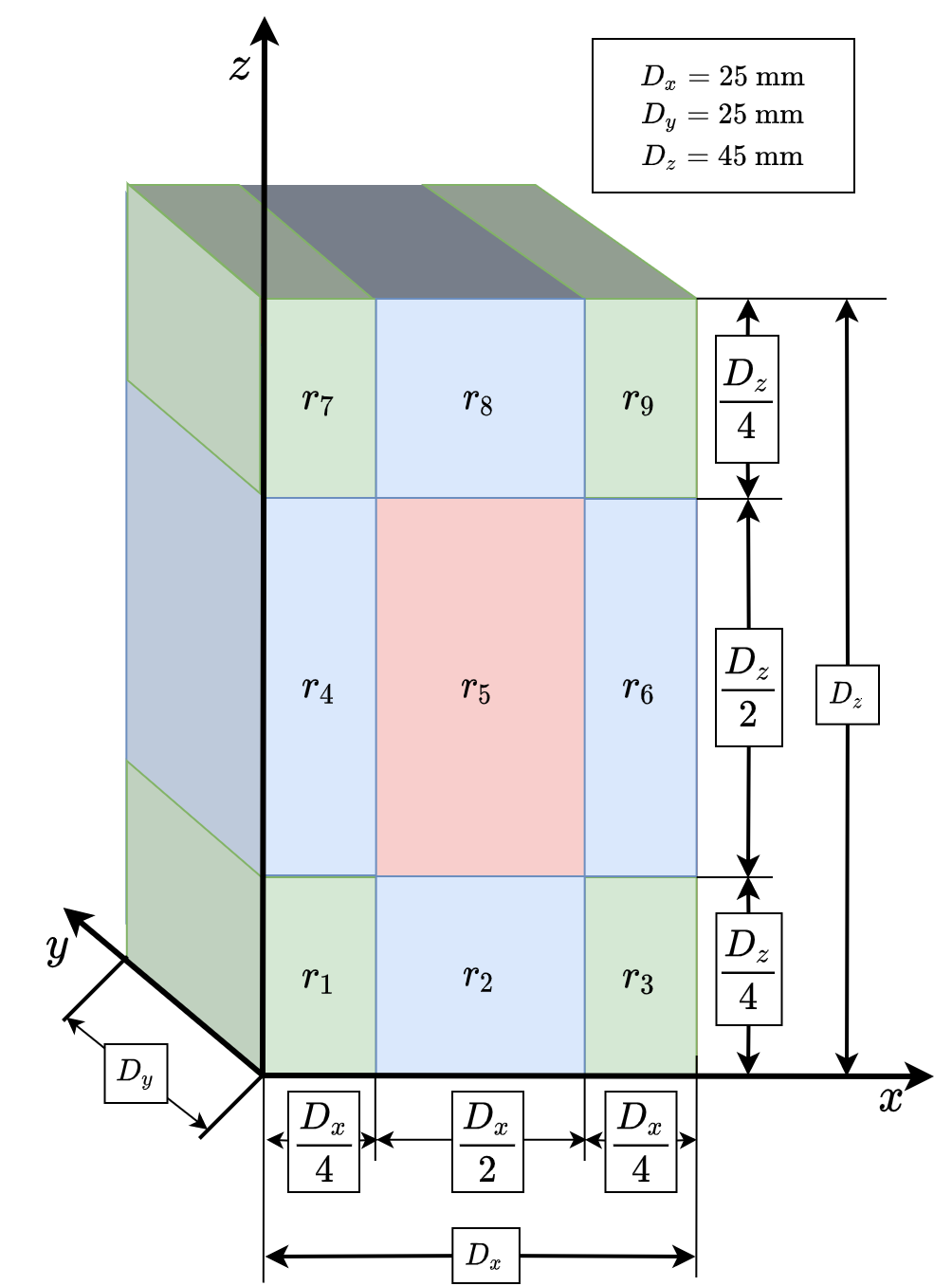}
\caption{The various regions  wherein the bulk packing fraction is measured. $D_x$,$D_y$ and $D_z$ are the dimensions of the box in the x and z directions respectively.}
\label{Fig:regions}
\end{center}
\end{figure}

\begin{itemize}
    \item Planar packing fraction variation along the x-direction and y-direction throughout the domain.
    \item Packing fraction variation along the z-direction throughout the domain along with the average bulk packing fraction $\eta_{a}$.
\end{itemize}
    For the cubic superquadric particles, we have also shown the following results as a demonstration of the kind of post processing possible from the voxel data.
\begin{itemize}
    \item Packing Fraction variation through the domain projected onto the xy-plane when seen as a heat-map from the z-direction.
    \item Packing Fraction variation through the domain projected onto the xz-plane when seen as a heat-map from the y-direction.
    \item Bulk packing fraction in regions of the domain represented by cuboids which span through the domain in the y-direction and divide the domain into 9 segments on the zx-plane when seen from the y-direction, as shown in \cref{Fig:regions}.
    \label{item: posts}
\end{itemize}
Also in a subsequent section, we utilise the ability to calculate the planar packing fraction in any arbitrary plane~(as discussed in \cref{sec:Calculation of Planar Packing Fraction along any Arbitrary Plane}) to compute the planar packing fraction variation along the face diagonal of the domain represented by the vector $\vec{n} = (1,1,0)$.
\subsection{Results and discussion}
\begin{figure*}[htbp]
    \centering
    \begin{subfigure}[t]{8.6 cm}
    \centering
    \includegraphics[width=0.9\linewidth]{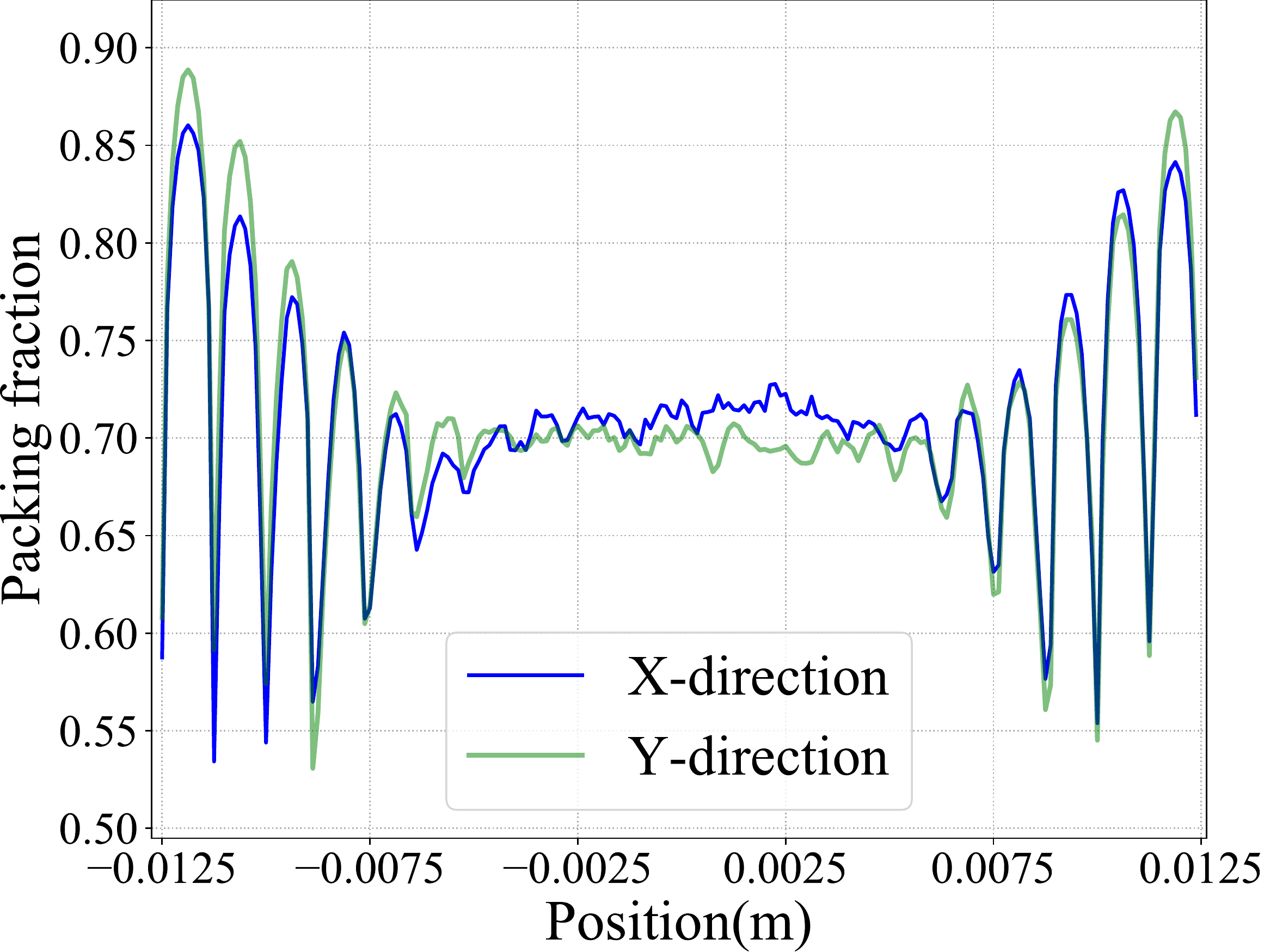}
    \caption{}
    \label{fig: sq_cube_xy}
    \end{subfigure}%
    \begin{subfigure}[t]{8.6 cm}
    \centering
    \includegraphics[width=0.9\linewidth]{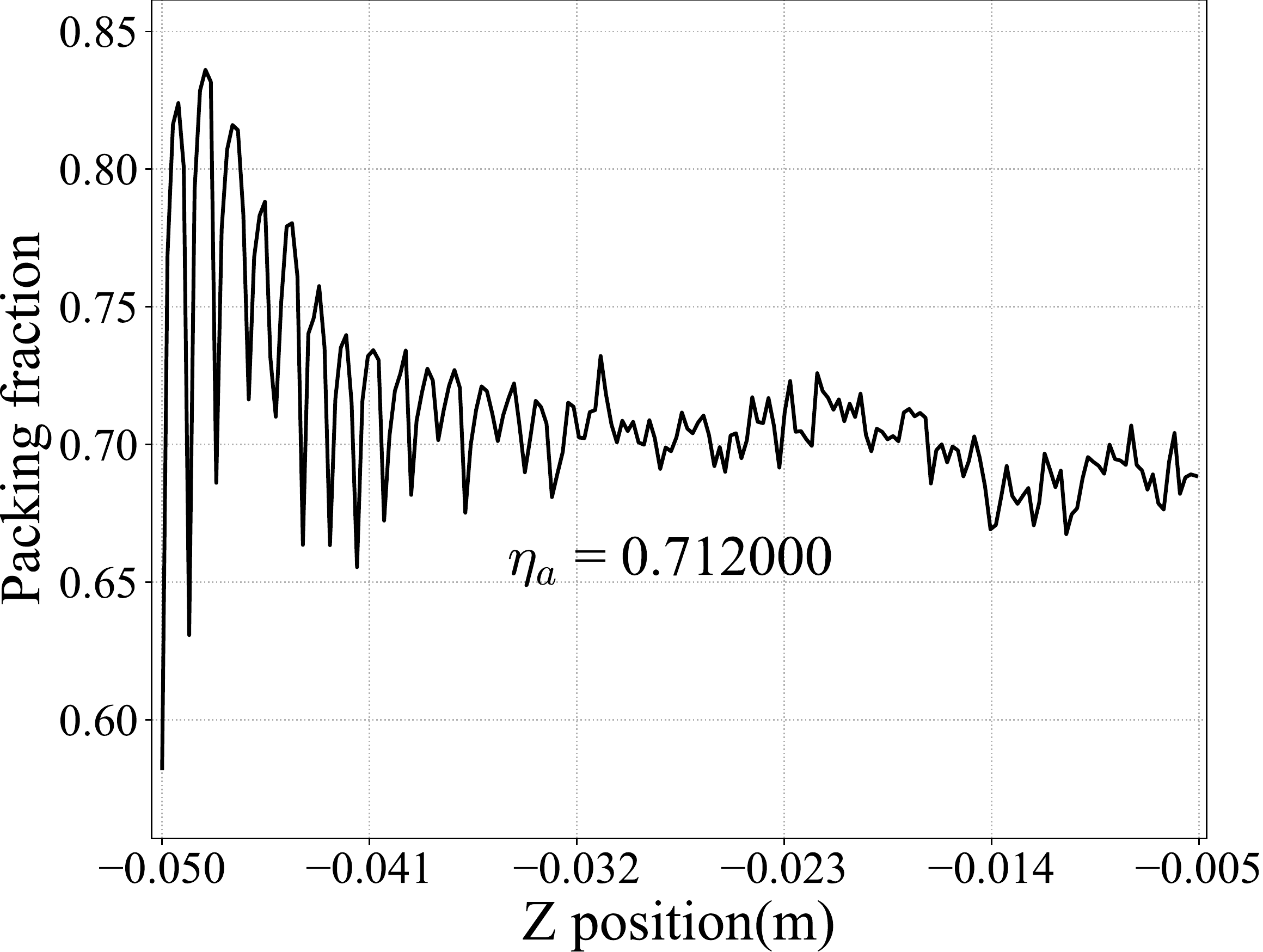}
    \caption{}
    \label{fig: sq_cube_z}
    \end{subfigure}
    \begin{subfigure}[t]{7.6208cm}
    \centering
    \includegraphics[width=0.9\linewidth]{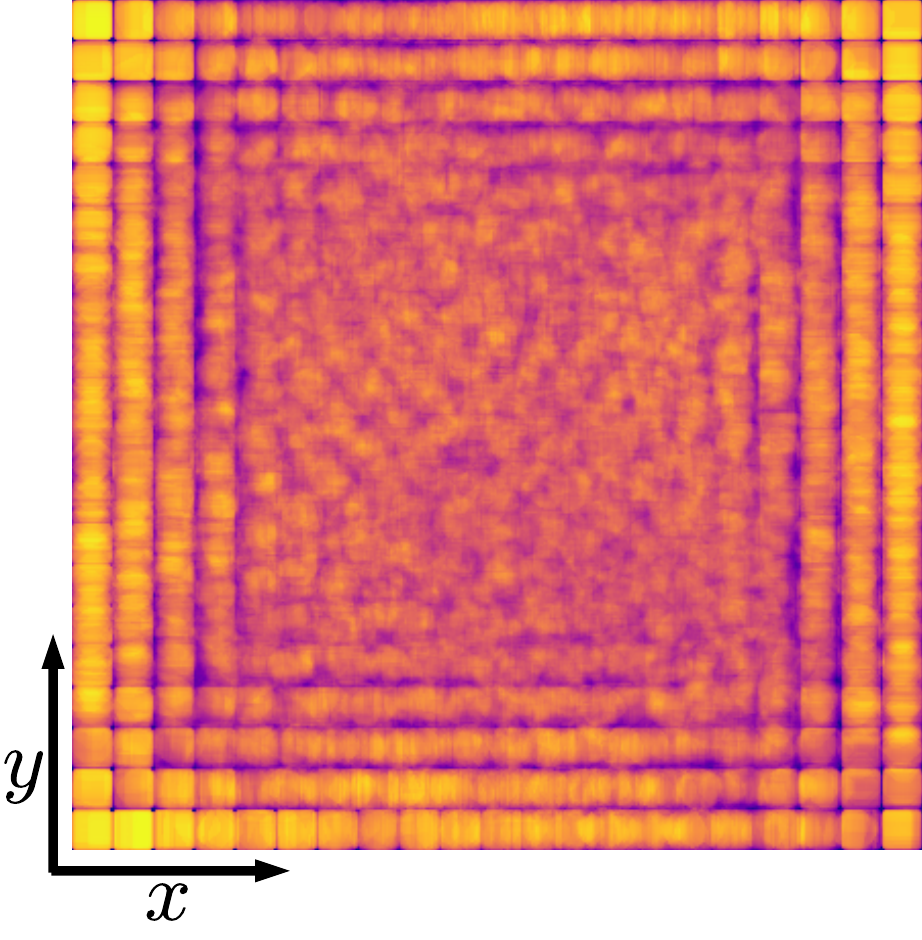}
    \caption{}
    \label{fig: sq_cube_z_surf}
    \end{subfigure}%
    \begin{subfigure}[t]{4.28 cm}
    \centering
    \includegraphics[width=0.9\linewidth]{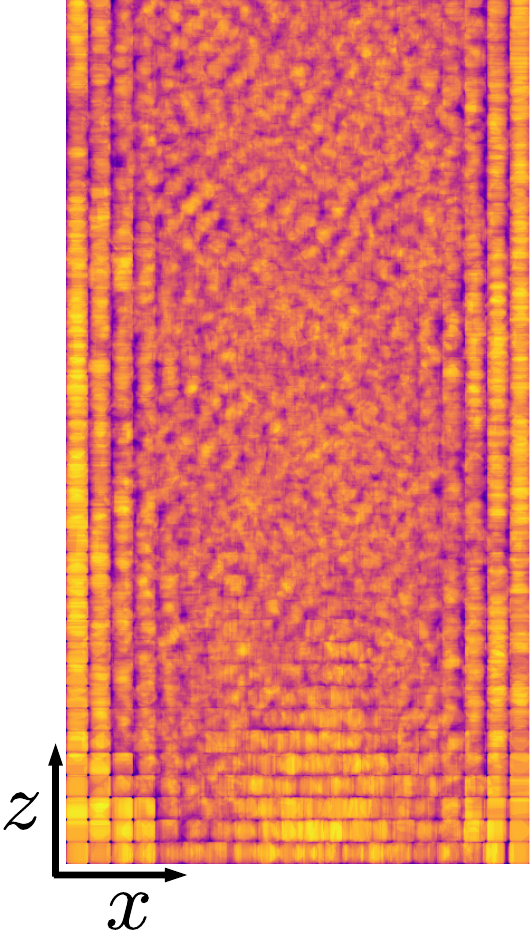}
    \caption{}
    \label{fig: sq_cube_y_surf}
    \end{subfigure}%
    \begin{subfigure}[t]{4.28 cm}
    \centering
    \includegraphics[width=0.9\linewidth]{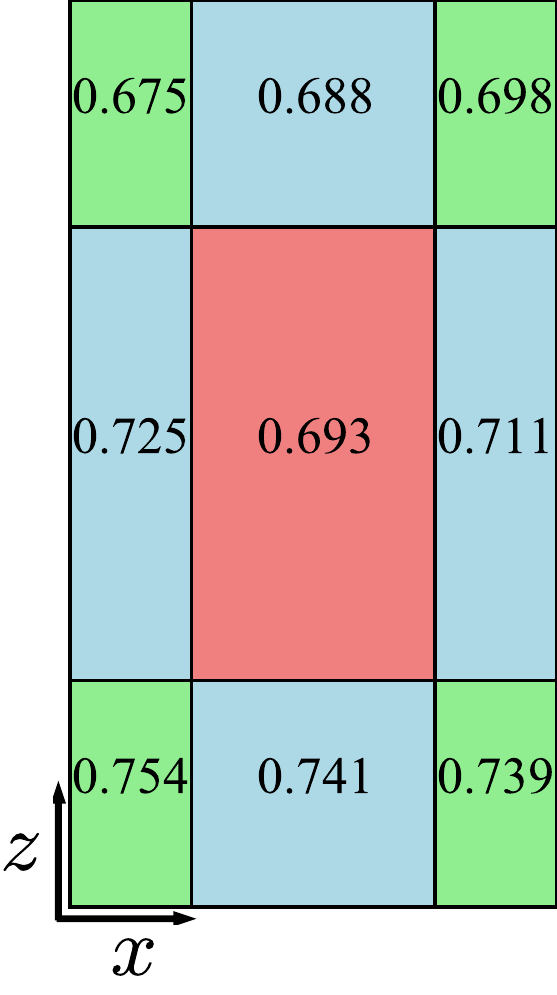}
    \caption{}
    \label{fig: sq_cube_reg_pf}
    \end{subfigure} 
    \caption{Various post processing results discussed in \cref{item: posts} for Case-1, i.e, cubic superquadric particles. (a) Packing fraction along the x-direction and y-direction,(b) Packing fraction along the z-direction;  $\eta_{a}=0.712$ represents the average packing fraction,(c) Heat-map of the packing fraction through the domain projected on to the xy-plane(lighter regions are densely packed and dark regions are loosely packed), (d)  Heat-map of the packing fraction through the domain projected on to the xz-plane(lighter regions are densely packed and dark regions are loosely packed), (e) bulk packing fraction  in the zones shown in \cref{Fig:regions}.}
    \label{Fig: Res_sq_cube}
\end{figure*}
For the cubic particles, a strong wall effect can be observed from the packing fraction variation plots~(see \cref{fig: sq_cube_xy}, \cref{fig: sq_cube_z}). The flat surfaces of the particles stack well against the flat walls and on top of each other allowing for a very tight packing as can be seen in \cref{fig: sq_cube_z_surf} and \cref{fig: sq_cube_y_surf}, wherein a clear stacking behavior is observed near the walls. This wall effect persists for 5 layers of particles for side walls and for nearly 12 layers for the bottom wall. The weight of the particles act as a compaction force on the particles closer to the bottom wall. This additional compaction forces the bottom particles to stack more densely, aiding in the deeper persistence of the wall effect of the bottom wall. This effect can also be observed in  \cref{fig: sq_cube_reg_pf} where regions near to the walls have a bulk packing fraction of greater than 0.7 whereas the centre region has a bulk packing fraction less than 0.7. We see that the average packing fraction is 0.712, which is quiet high as the particle shape and the container shape are compatible with each other. 
\begin{figure*}[htbp]
    \centering
    \begin{subfigure}[t]{8.6 cm}
    \centering
    \includegraphics[width=0.9\linewidth]{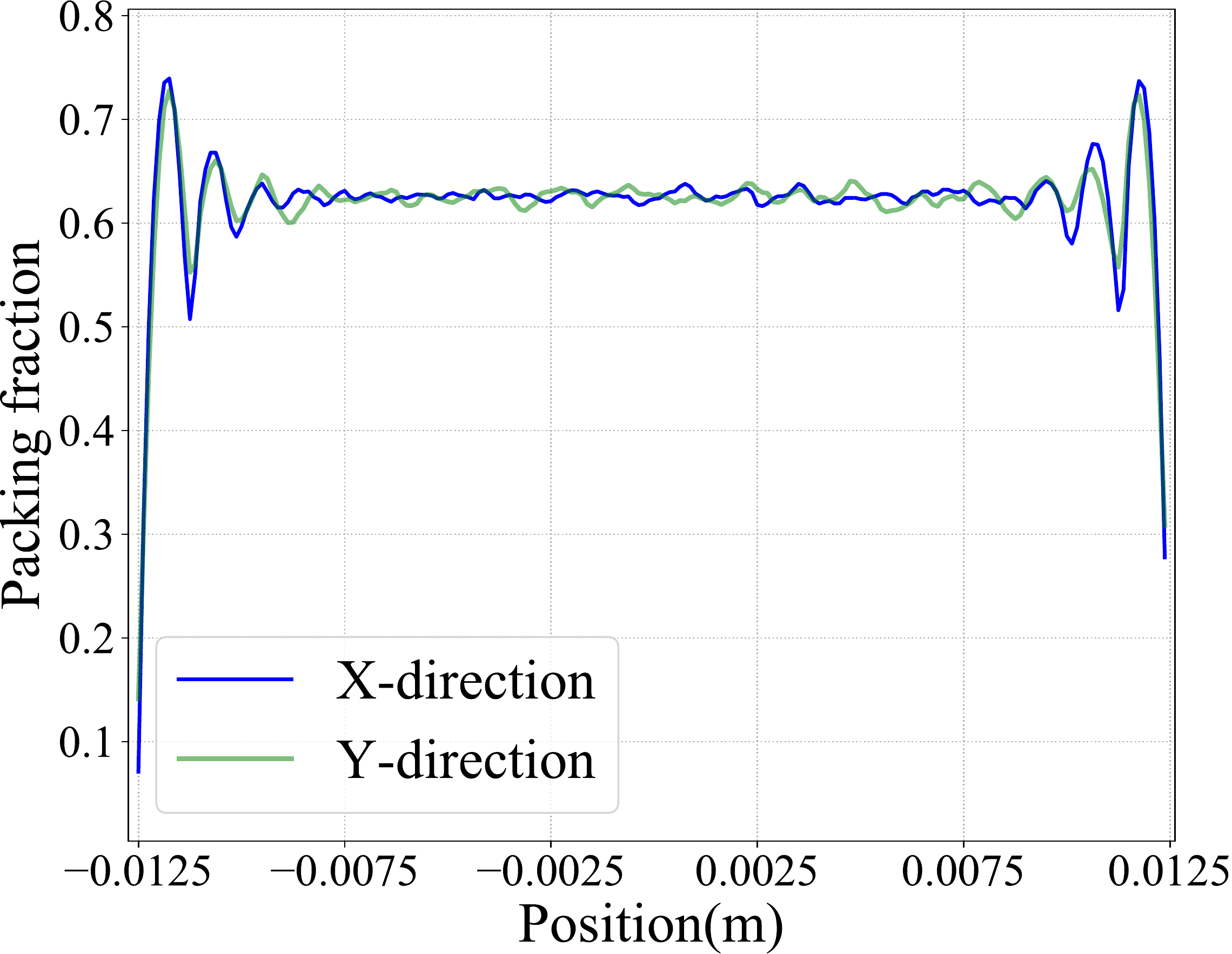}
    \caption{}
    \label{fig: sq_elp_xy}
    \end{subfigure}%
    \centering
    \begin{subfigure}[t]{8.6 cm}
    \centering
    \includegraphics[width=0.9\linewidth]{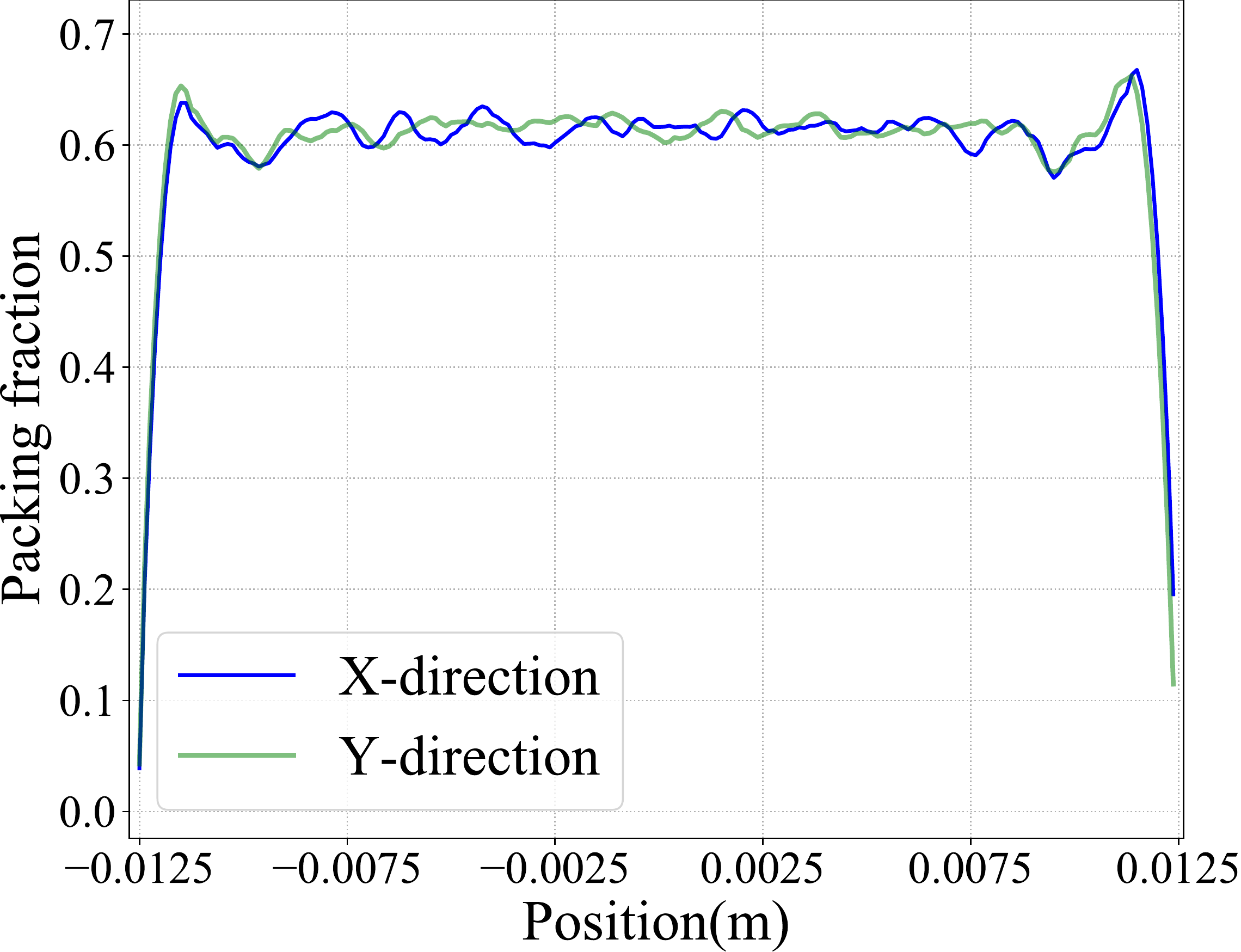}
    \caption{}
    \label{fig: ms_coffee_xy}
    \end{subfigure}
    \begin{subfigure}[t]{8.6 cm}
    \centering
    \includegraphics[width=0.9\linewidth]{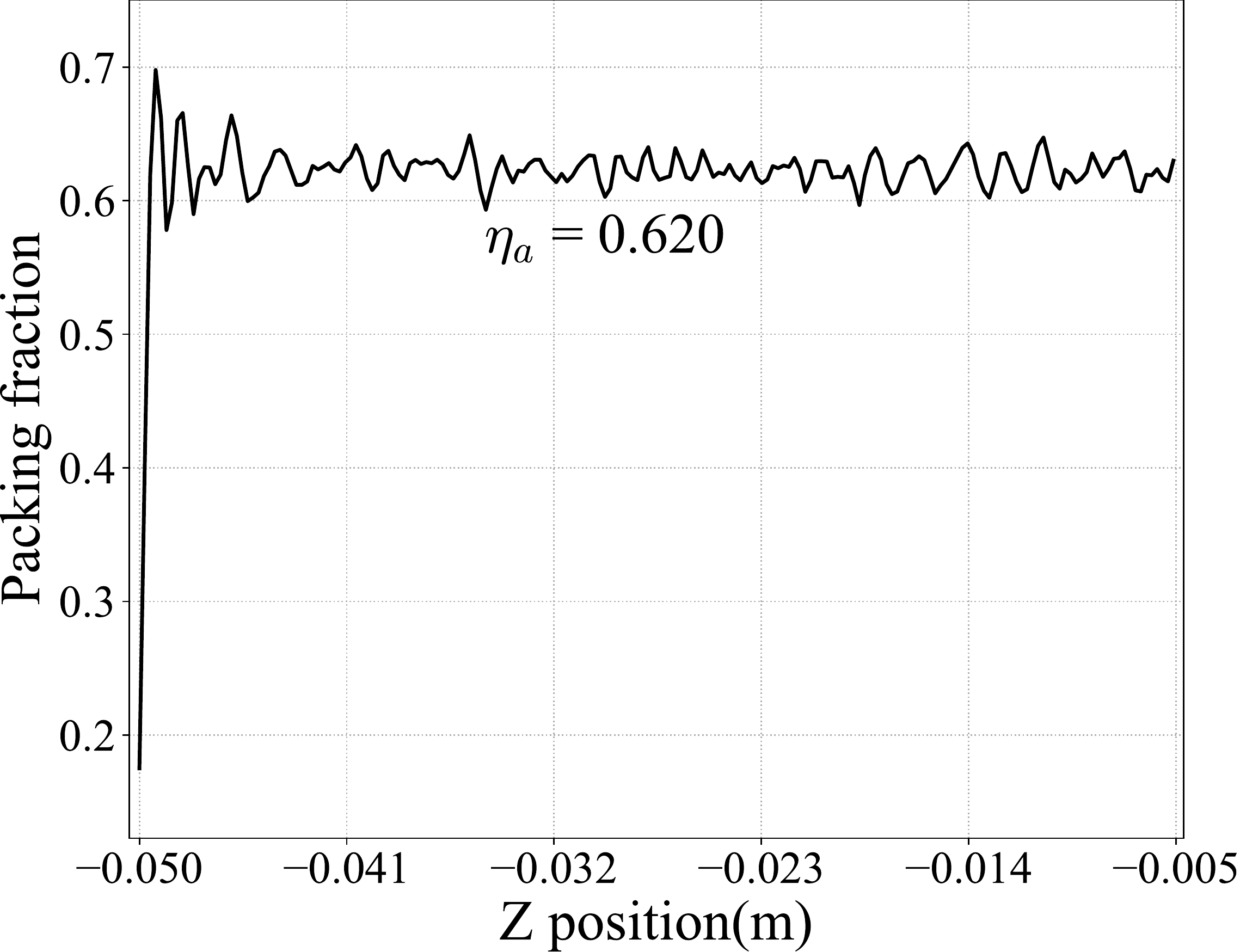}
    \caption{}
    \label{fig: sq_elp_z}
    \end{subfigure}%
    \begin{subfigure}[t]{8.6 cm}
    \centering
    \includegraphics[width=0.9\linewidth]{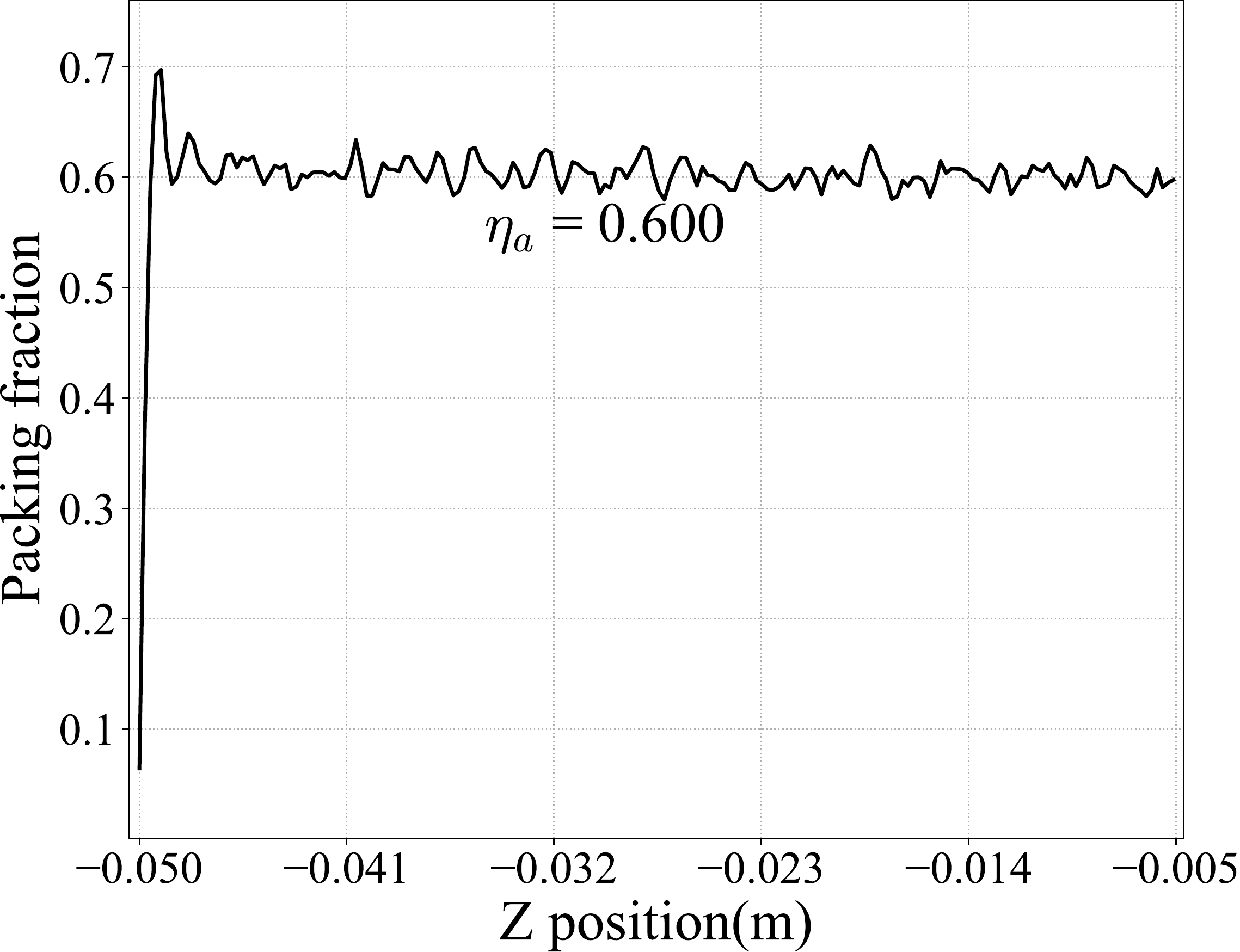}
    \caption{}
    \label{fig: ms_coffee_z}
    \end{subfigure}
    \caption{Packing fraction along the x-direction and y-direction for (a) superquadric ellipsoidal particle assembly and (b) multisphere coffee bean particle assembly. Packing fraction along the z-direction (c) for superquadric ellipsoidal particle assembly and (d) for multisphere coffee bean particle assembly.}
    \label{Fig: Res_ms_coffee}
\end{figure*}
For the superquadric ellipsoidal particles we see a weak wall effect when compared to the cubic particles, We can observe two prominent peaks close to the side walls as seen in \cref{fig: sq_elp_xy}. For multisphere coffee bean shaped particles, we don't see any significant wall effect, as can be seen in \cref{fig: ms_coffee_xy}. The packing fraction is nearly flat throughout the domain in both the x and y directions. The packing fraction for both the ellipsoidal and coffee bean particle is close to 0.1 at the extremes of the domain (closest to the wall), indicating the poor mating of the curved surfaces of the particles against the the flat surface of the wall. The same cannot be said for the cubic particles whose packing closest to the walls starts at 0.6 and quickly increases from there. 
The packing in z direction for both the ellipsoidal particle and the coffee bean shaped particle shows similar trend. The packing fraction starts close to 0.1 and quickly increases and converges near the average packing fraction (see \cref{fig: sq_elp_z,fig: ms_coffee_z}). The average packing fraction of the coffee bean shaped particle assembly is $0.6$ and for the ellipsoidal particle assembly, it is $0.62$, indicating similar packing densities for both particles. The heat-maps and regional packing for the ellipsoidal particles and the coffee beans can be found in the supplementary material(Figure S1 and S2).
\subsection{Packing fraction in an arbitrary direction}
\label{subsec:Packing fraction in arbitrary direction}
\begin{figure*}[htbp]
    \centering
    \begin{subfigure}[t]{7.74 cm}
    \includegraphics[width=0.99\textwidth]{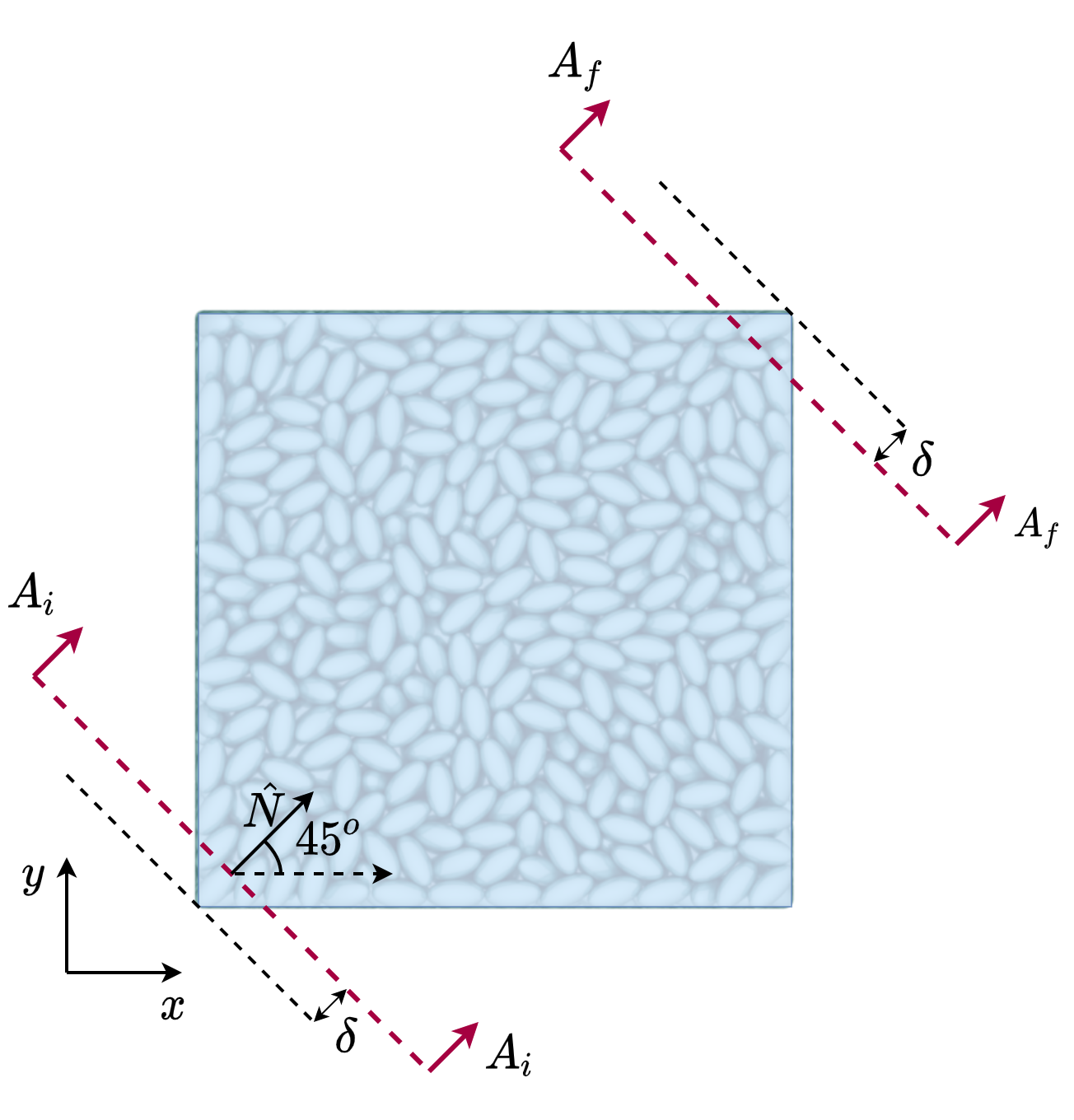}
    \caption{} 
    \label{Fig:diag_plane}
    \end{subfigure}
    \centering
    \begin{subfigure}[t]{8.6 cm}%
    \centering
    \includegraphics[width=0.9\linewidth]{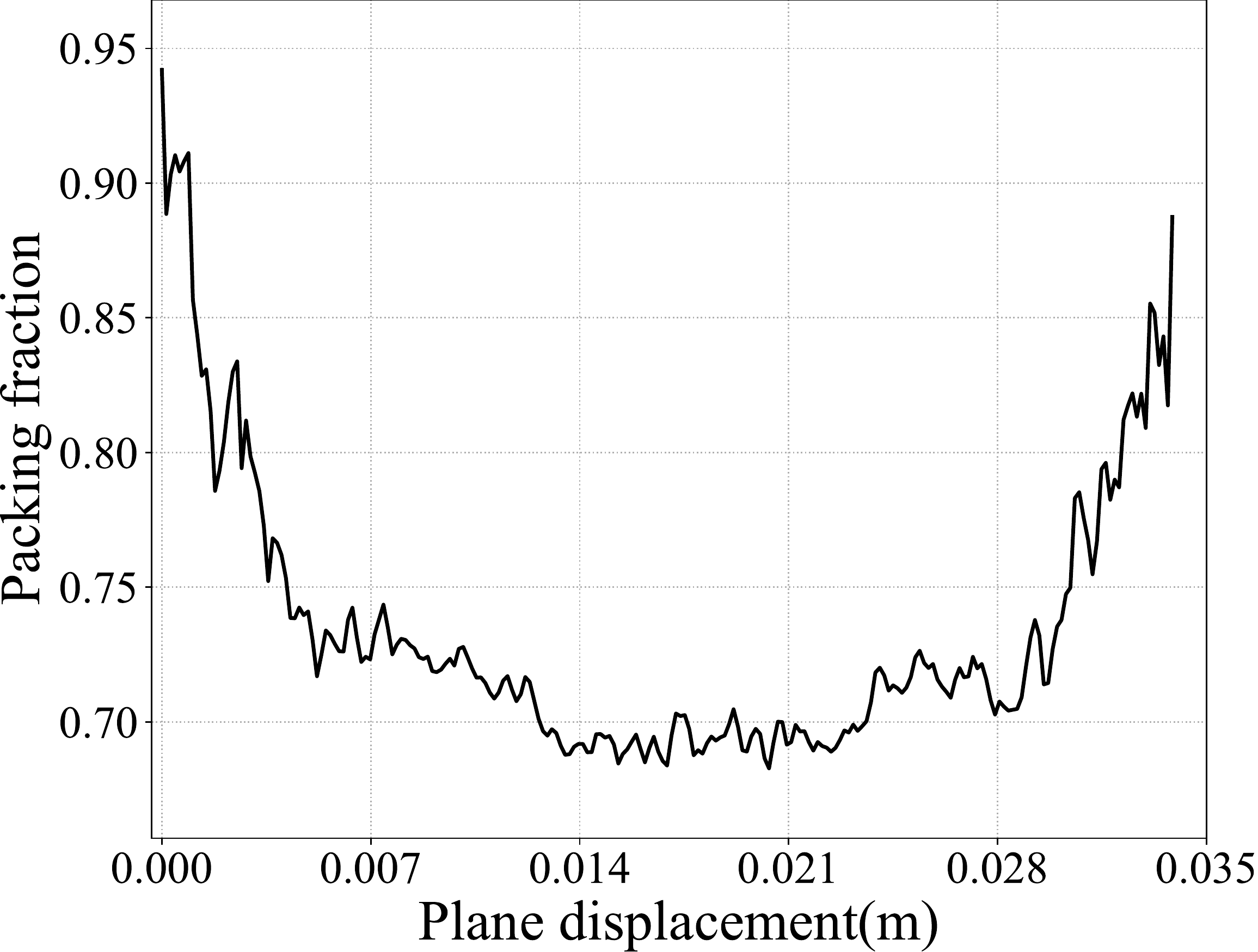}
    \caption{}
    \label{Fig:diag_plane_sq_cube}
    \end{subfigure}
    \begin{subfigure}[t]{8.6 cm}
    \centering
    \includegraphics[width=0.9\linewidth]{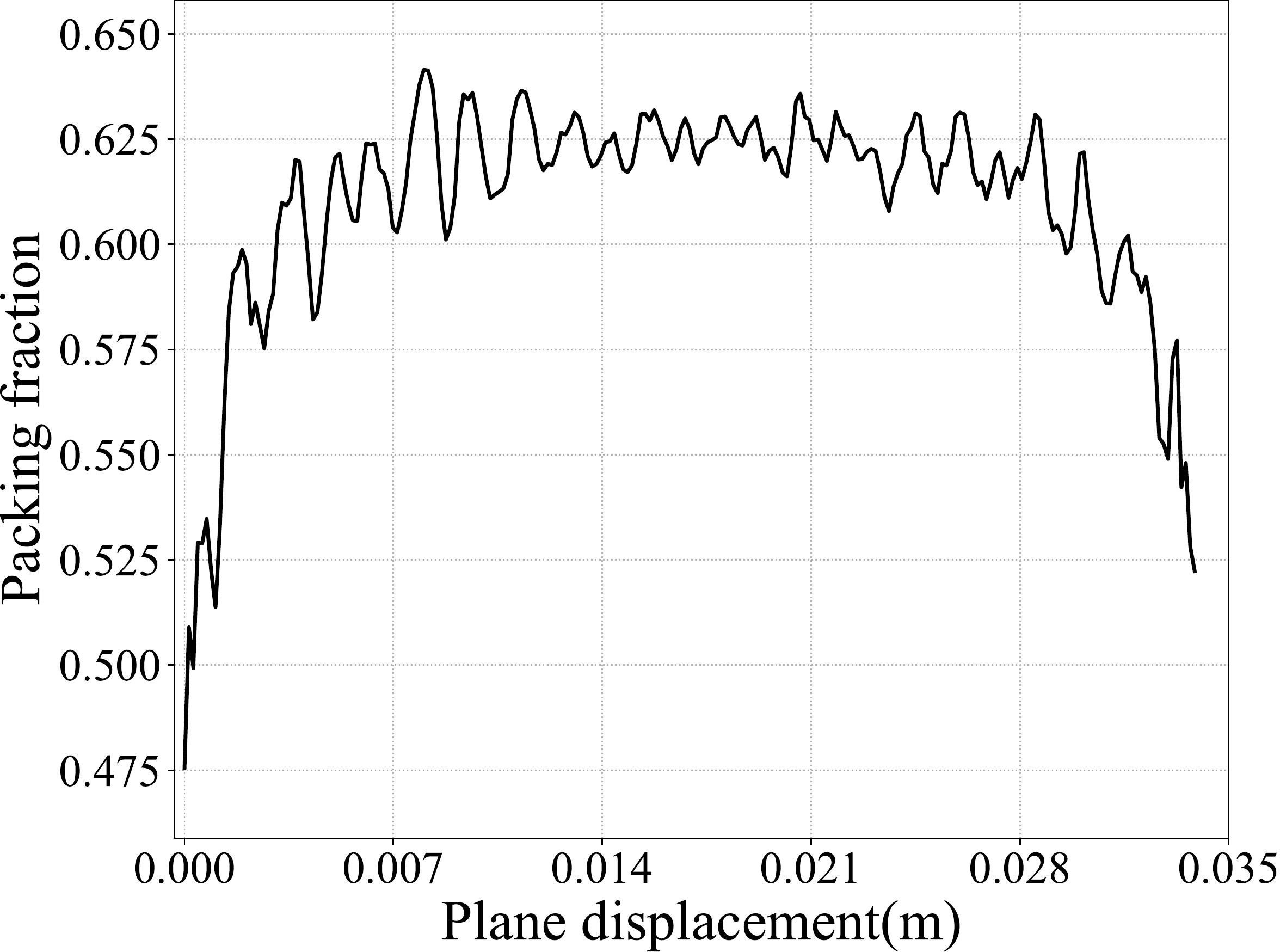}
    \caption{}
    \label{Fig:diag_plane_sq_elp}
    \end{subfigure}%
    \begin{subfigure}[t]{8.6 cm}
    \centering
    \includegraphics[width=0.9\linewidth]{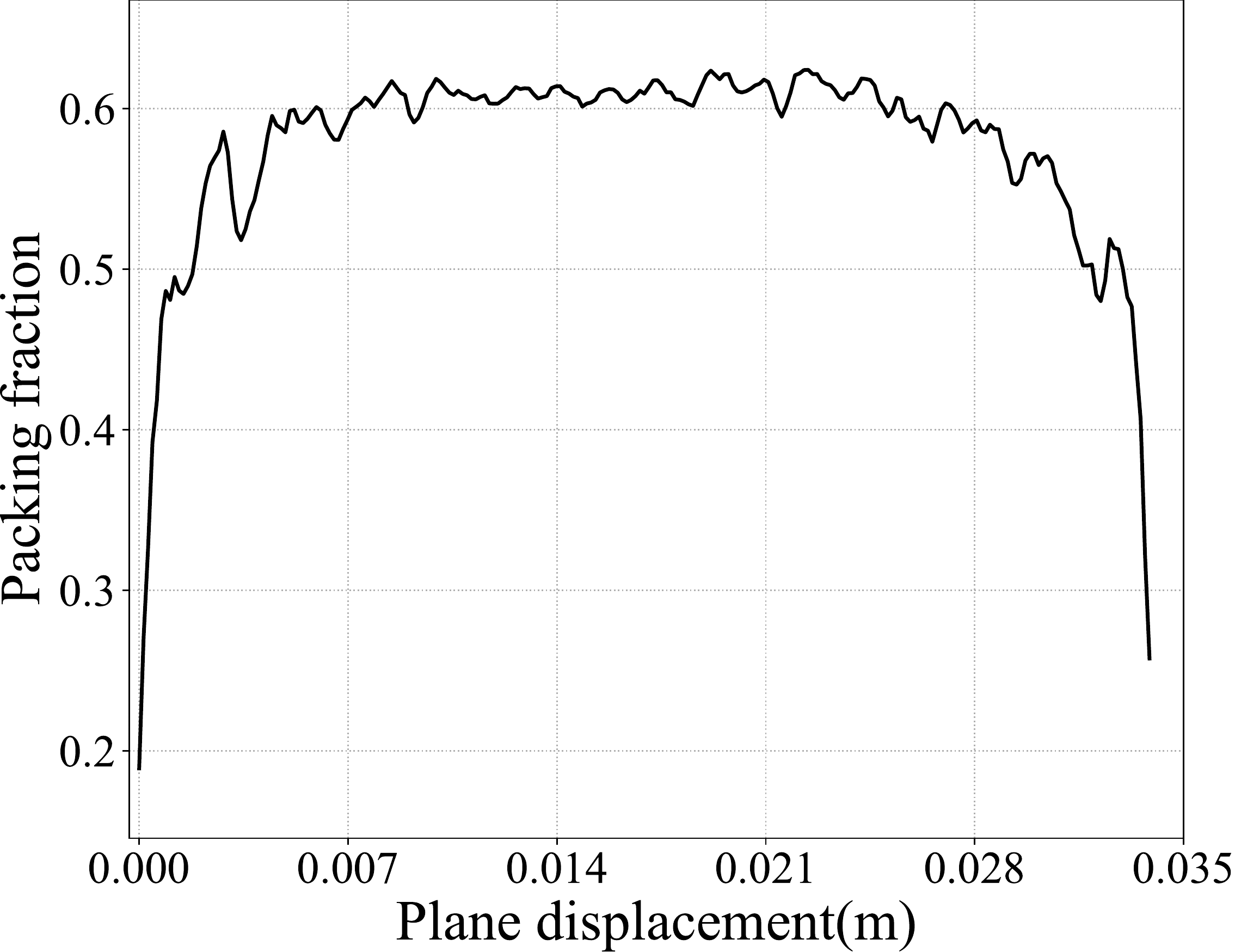}
    \caption{}
    \label{Fig:diag_plane_ms_coffee}
    \end{subfigure}
    \caption{(a) The direction $\hat{N}$, along with the planes on either extremes $A_i$ and $A_f$ which are at a offset of $\delta = \SI{0.7}{mm}$(exaggerated in the diagram) from either extremes, the packing fraction variation is calculated along $\hat{N}$ from $A_i$ to $A_f$ for (b) cubic superquadric particle (c) ellipsoidal superquadric particle (d) coffee bean shape multisphere particle.} 
    \label{Fig: diagonal packing fraction}
\end{figure*}

In this subsection, we demonstrate the computation of packing fraction variation in a plane oriented at any arbitrary direction $\hat{N}$ as shown \cref{Fig:diag_plane}, from the plane $A_i$ to the plane $A_f$ with an offset of $\delta = \SI{0.7}{mm}$ on either extremes. The blue square with the particles inside represents the container as seen from a top view.
For the cubic superquadric particles~(see \cref{Fig:diag_plane_sq_cube}), a strong densification (local packing fraction $\approx 0.9$) near the corners is observed. This behaviour can be explained from \cref{fig: sq_cube_z_surf}, where we can observe very strong stacking near the corners and a more randomised packing near the centre causes a reduction in packing fraction. The packing fraction gradually reduces as we move away from the corners, reaching a minimum value close to 0.7 at  the middle of the domain.
For both the superquadric ellipsoidal~(see \cref{Fig:diag_plane_sq_elp}) and multisphere coffee beans~(see \cref{Fig:diag_plane_ms_coffee}), we see similar packing variation. We observe that the packing fraction close to the corners is much lower and the values plateau close to their respective average packing fraction as we move away from the corners. This is due to the curvature of the particles not allowing them to fit properly near the corner of the cube, resulting in gaps at the corners, reducing the packing fraction near these regions. We also notice that the packing fraction for the coffee beans is a little more uniform (flat) in the middle when compared to the ellipsoidal particles. This may be because the coffee bean is a more irregular shape when compared to the ellipsoidal particle, which causes a more homogeneous distribution of particles or/inversely voids in the domain, resulting in a uniform packing fraction. Where as the ellipsoidal particles spontaneously form more orderly packing structures, which in-turn may cause some periodic behaviour in the packing fraction variation.
%=================================================%
\clearpage
\section{Conclusion}
In this paper, we have introduced a voxelization based post processing workflow to analyse the packing structure of non-spherical particle assembles obtained from DEM simulations of either superquadric or multisphere particles. We have also provided an algorithm to compute the planar packing fraction on an arbitrary plane. Finally, we have used these algorithms to obtain packing characteristics of assemblies of simple non-spherical particles. We have seen a strong stacking behaviour of the cubic superquadric particles, where as ellipsoidal superquadric and coffee bean shaped multisphere particles showed a nearly randomised, packing with very minimal wall effect.
The algorithm presented in this paper allows us to study packing of non-spherical particles in a unique way. We can isolate specific regions to study the packing in them, scan the assembly to generate packing heat-maps, and characterise packing fraction variation of non-spherical particles in any arbitrary direction. Most granular material seen in real life are not spherical and deviation from the spherical shape alters the packing in drastic ways, hence, this algorithm would prove useful for designing systems with arbitrary shaped particles found in several real-world granular materials.
\section*{acknowledgements}
The authors acknowledge the Corporate Social Responsibility (CSR) grant and the EDEM software license provided by Altair India Pvt Limited.
%
% \balance
% \section*{References}
% \bibliographystyle{unsrtnat}
\bibliography{references.bib}

\end{document}

\begin{center}
\textbf{ \Large \underline{Supplementary Information}\\
\vspace{0.3 cm}
Voxelization based packing analysis for discrete element simulations of non-spherical particles}\\
\vspace{0.2 cm}
\noindent{\large Venkata Rama Manoj Pola and Ratna Kumar Annabattula}\\
\vspace{0.2 cm}
{\small Department of Mechanical Engineering, Indian Institute of Technology Madras, Chennai - 600036, India}\\
{\small Additive Manufacturing Group - Centre of Excellence in Materials and Manufacturing for Futuristic Mobility, Indian Institute of Technology Madras, Chennai - 600036, India}\\
\vspace{0.2cm}
\noindent{\large Raghuram Karthik Desu}\\
{\small Department of Mechanical Engineering, National Institute of Technology, Tiruchirappalli - 620015, India}
\end{center}

\vspace{1 cm}

\section{Additional post processing results of superquadric ellipsoidal particle}
\begin{figure*}[htbp]
    \begin{subfigure}[t]{7.6208cm}
    \centering
    \includegraphics[width=0.9\linewidth]{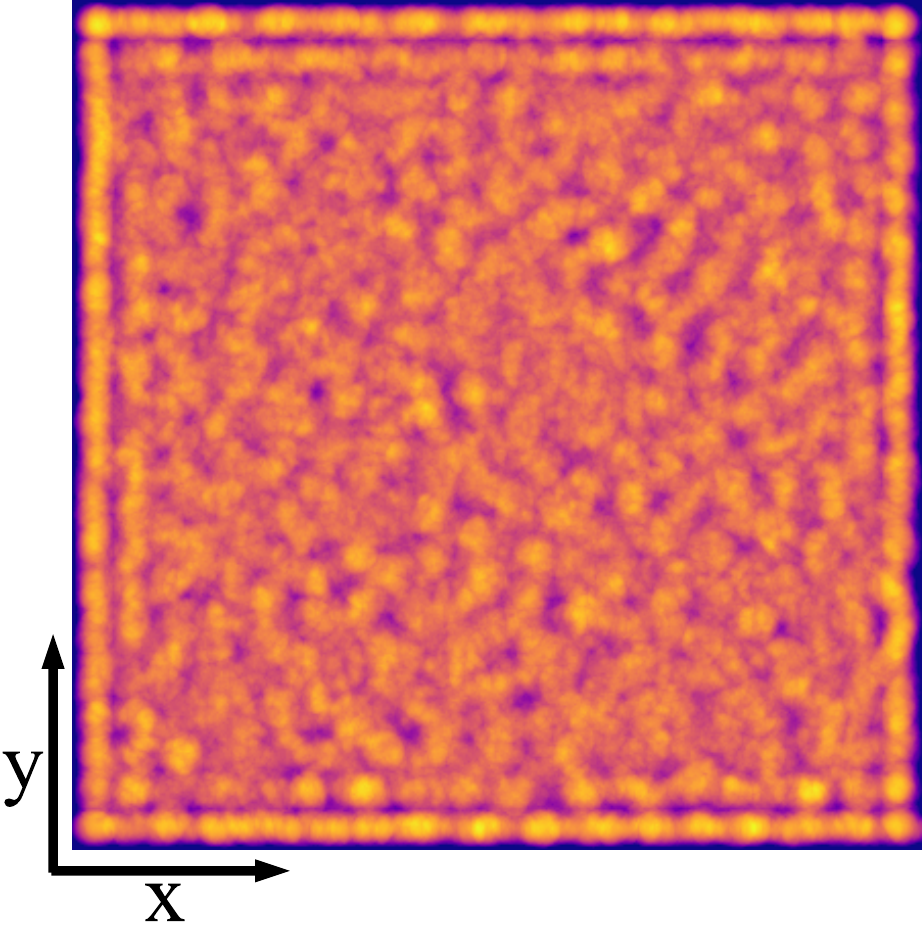}
    \caption{}
    \label{fig: sq_elp_z_surf}
    \end{subfigure}%
    \begin{subfigure}[t]{4.28 cm}
    \centering
    \includegraphics[width=0.9\linewidth]{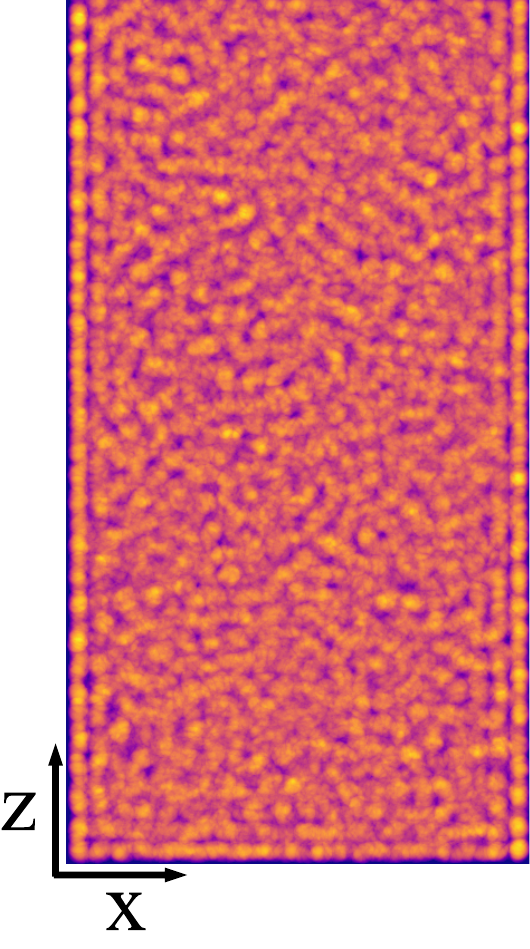}
    \caption{}
    \label{fig: sq_elp_y_surf}
    \end{subfigure}%
    \begin{subfigure}[t]{4.28 cm}
    \centering
    \includegraphics[width=0.9\linewidth]{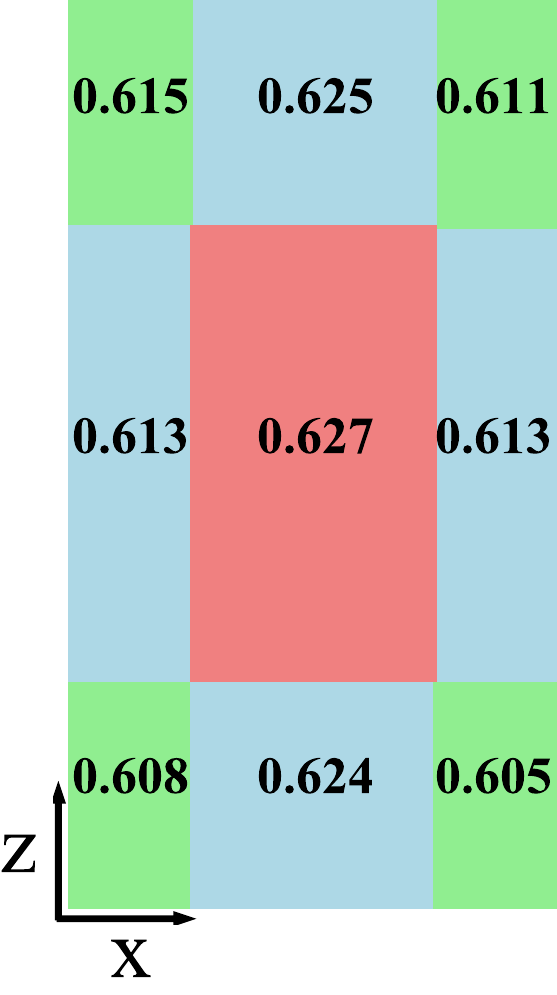}
    \caption{}
    \label{fig: sq_elp_reg_pf}
    \end{subfigure} 
    \caption{(a) Heat-map of the packing fraction through the domain projected on to the xy-plane(lighter regions are densely packed and dark regions are loosely packed), (b)  Heat-map of the packing fraction through the domain projected on to the xz-plane(lighter regions are densely packed and dark regions are loosely packed), (c) bulk packing fraction  in the zones shown in Figure 11.}
    \label{Fig: Res_sq_elp}
\end{figure*}
\newpage
\section{Additional post processing results of multisphere coffee bean shpaed particles}
\begin{figure*}[htbp]
    \begin{subfigure}[t]{7.6208cm}
    \centering
    \includegraphics[width=0.9\linewidth]{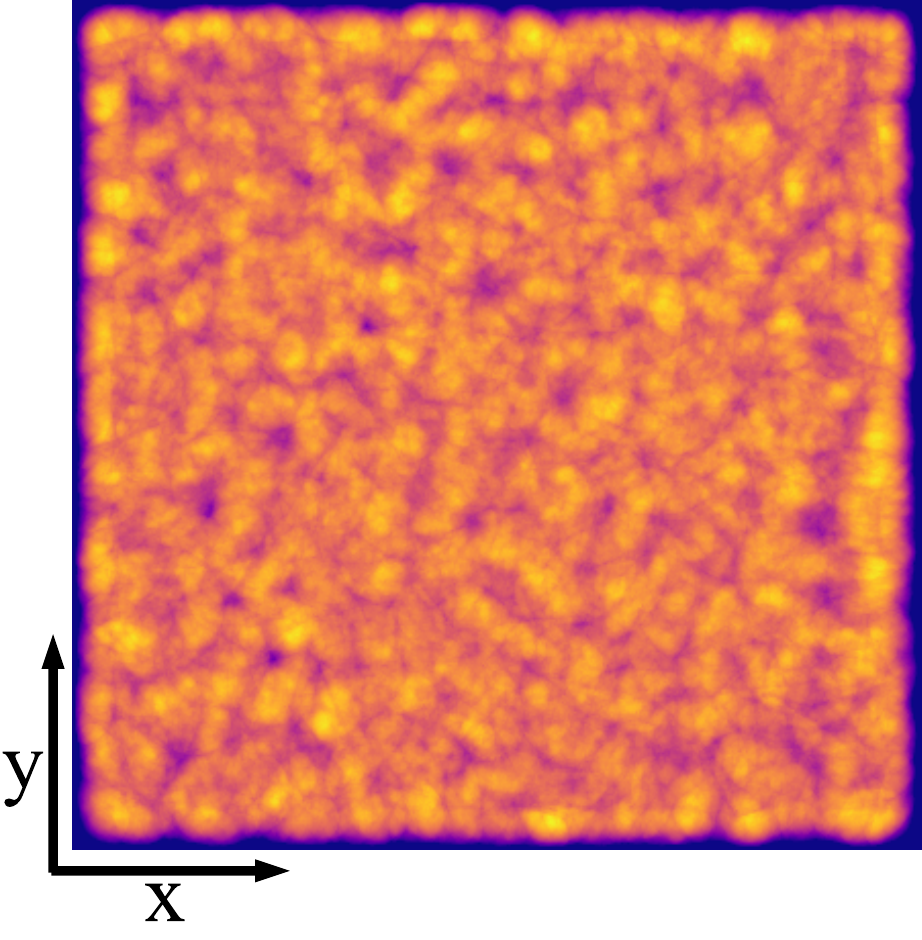}
    \caption{}
    \label{fig: ms_coffee_z_surf}
    \end{subfigure}%
    \begin{subfigure}[t]{4.28 cm}
    \centering
    \includegraphics[width=0.9\linewidth]{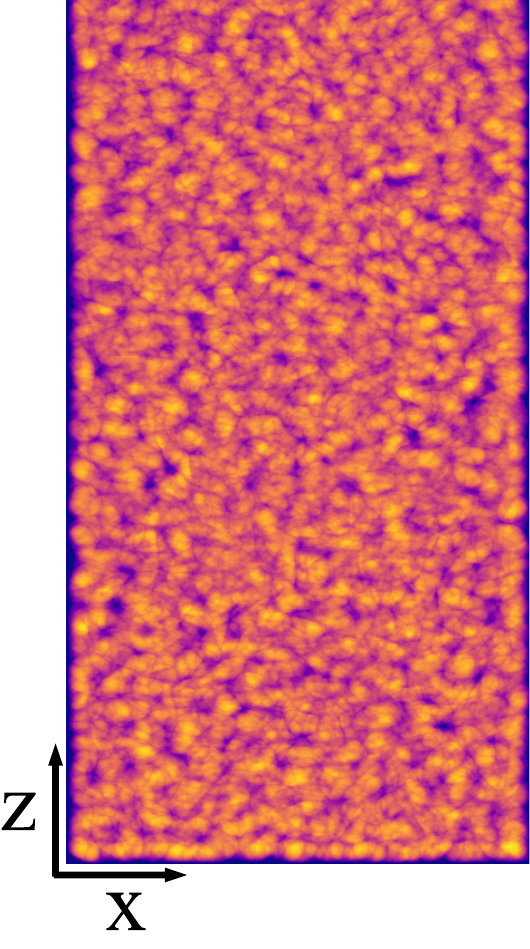}
    \caption{}
    \label{fig: ms_coffee_y_surf}
    \end{subfigure}%
    \begin{subfigure}[t]{4.28 cm}
    \centering
    \includegraphics[width=0.9\linewidth]{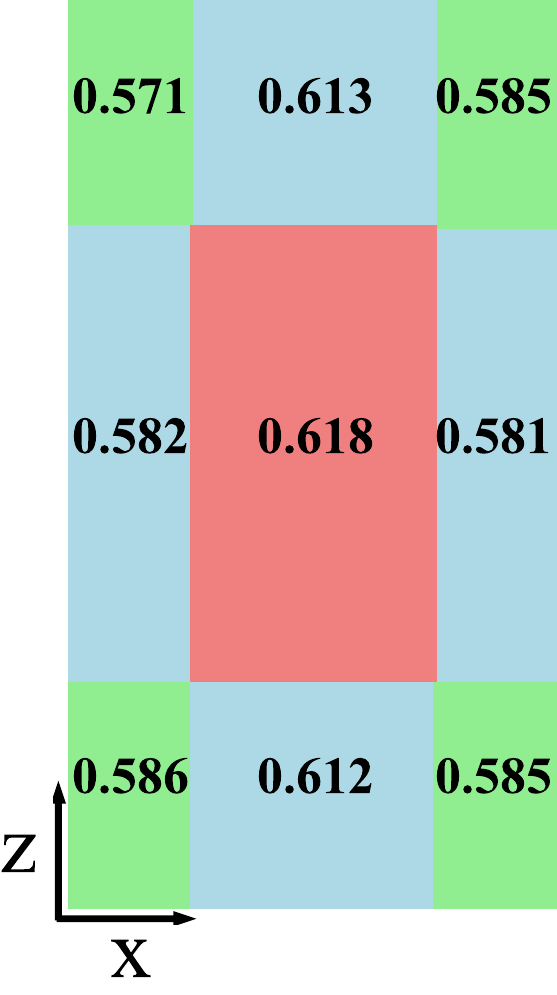}
    \caption{}
    \label{fig: ms_coffee_reg_pf}
    \end{subfigure} 
    \caption{(a) Heat-map of the packing fraction through the domain projected on to the xy-plane(lighter regions are densely packed and dark regions are loosely packed), (b)  Heat-map of the packing fraction through the domain projected on to the xz-plane(lighter regions are densely packed and dark regions are loosely packed), (c) bulk packing fraction  in the zones shown in Figure 11.}
    \label{Fig: Res_ms_coffee}
\end{figure*}
For the ellipsoidal particles, we can observe a small wall effect for both the side and bottom wall(see \cref{fig: sq_elp_z_surf,fig: sq_elp_y_surf}, which agrees with the packing fraction variation plots(see Figure 13a and 13c). For the coffee bean particles, we cannot observe any significant wall effect(see \cref{fig: ms_coffee_z_surf,fig: ms_coffee_y_surf}). For either cases, we see the the packing fraction closest to the walls is very low, which is also observed in the packing fraction variation plots(see Figure 13).From the regional packing fraction of both the particle shapes(see \cref{fig: sq_elp_reg_pf} and \cref{fig: ms_coffee_reg_pf}, we see that the regions closest to the walls have lower packing than the regions further away from walls. This implies that flat walls tend to reduce the packing density of granular assemblies with these particle shapes.